\documentclass[]{article}
\usepackage{amsmath}
\usepackage{amssymb}
\usepackage{graphicx}
\usepackage{float}
\usepackage{color}
\usepackage{geometry}
\usepackage[numbers]{natbib}
 
\everymath{\displaystyle}
\newcommand{\pd}[2]{\displaystyle \frac{ \partial #1}{ \partial #2}}
\newcommand{\sgn}[1]{\mathrm{sgn}(#1)}

\def\ie{{\sl i.e.}}

\title{Hyperbolicity of the modulation equations for the Serre-Green-Naghdi model} 
{\author{Sergey Tkachenko
     \thanks{Aix Marseille Univ, CNRS, IUSTI,   UMR 7343,  Marseille, France,   sergey.tkachenko@univ-amu.fr}, 
        Sergey Gavrilyuk
     \thanks{Corresponding author: Aix Marseille Univ, CNRS, IUSTI,   UMR 7343,  Marseille, France,   sergey.gavrilyuk@univ-amu.fr},  
	 Keh-Ming Shyue 
         \thanks{Institute of Applied Mathematical Sciences,
  		 National Taiwan University, Taipei 106, Taiwan, shyue@ntu.edu.tw}}
\begin{document}
\maketitle
\begin{abstract}

Serre-Green-Naghdi equations (SGN equations) is the most simple dispersive model of
long water waves having ``good" mathematical and physical properties. 
First, the model is a mathematically justified approximation of  the exact water wave problem. Second, the SGN equations are the Euler-Lagrange equations coming from Hamilton's principle of stationary action with a natural  approximate Lagrangian.  Finally, the equations are Galilean invariant which is necessary for physically relevant mathematical models. 
	
	We have  derived  the  modulation equations to the SGN model and show that 
they are  strictly hyperbolic for any wave amplitude, \ie, the periodic wave trains are modulationally stable. 
Numerical tests for the full SGN equations are shown. The results confirm the modulational stability analysis. 

\end{abstract}

\section{Introduction}
One usually uses two methods to obtain the modulation equations for a reversible dispersive system: 
the Whitham averaged Lagrangian method~\cite{Whitham_1974} or 
averaging of the corresponding conservation laws (cf.~\cite{Bhatnagar_1979,Kamchatnov_2000}). 
Both of them give the same system for slowly varying  wave train characteristics 
(wave length and amplitude, for example). 
If the corresponding system of modulation equations is hyperbolic (elliptic), 
one says that the corresponding wave trains are modulationally stable (unstable). 
The relation between the modulational instability and, for example,  
the classical spectral instability is not completely understood. 
An important result for a class of Hamiltonian systems  has been obtained in~\cite{Benzoni_2014}:  
the hyperbolicity of modulation  equations is {\it necessary} for the spectral stability 
of periodic traveling waves. 

The  modulational instability (wavetrain instability) is often generically called  
``Benjamin--Feir instability" even if, formally,   
this last instability concerns only about the surface gravity water waves  
(Benjamin and Feir~\cite{Benjamin_1967a} and Zakharov~\cite{Zakharov_1968} in the case of deep-water waves,  
and  Benjamin~\cite{Benjamin_1967b} in the case of finite-depth waves). 
We refer  interested readers  to the article~\cite{Zakharov_2009} where the history of the  
modulational instability theory is presented.  
In the case of small amplitudes, the hyperbolicity (ellipticity) condition can 
easily be formulated in terms of 
the non-linear amplitude dependent dispersion relation (see~\cite{Whitham_1974}, chapter~$15$). 
A recent application of such an approach can be found in~\cite{Meiden_2016} for a ``conduit" equation. 
However, the study  of hyperbolicity of the modulation equations in the case of 
{\it large amplitude solutions} is a more difficult problem. 
For integrable systems, the hyperbolicity of modulation equations and existence of the Riemann invariants 
were established, for example,  
for Korteweg-de Vries equation (KdV equation)~\cite{Whitham_1974}, 
for nonlinear Schr\"{o}dinger equation (NLS equation)~\cite{Pavlov_1987}, 
for sine-Gordon equation \cite{Forest_1983,  Gurevich_Gershenzon_1989}, 
for Benjamin-Ono equation \cite{Dobrokhotov_1991} (see also the book~\cite{Kamchatnov_2000} 
for further references). 
For non-integrable systems, one  can cite~\cite{Hur_2015} where the Whitham equation was studied 
and modulational instability for short enough waves was shown, 
or~\cite{Liapidevskii_Teshukov_2000} and~\cite{Johnson_2018} where the the regions of modulational 
and spectral stability for roll waves 
to the Saint-Venant equations were determined.  
In general,  a ``right choice"  of unknowns in which the modulation equations 
are written is necessary to have  explicit (or almost explicit) expressions of 
the  corresponding  characteristic values. 
Such a choice is not at all obvious. 

The aim of this work is to study the modulational stability of periodic waves to 
Serre-Green-Naghdi equations (SGN equations). 
One-dimensional SGN equations can be written in the Eulerian coordinates in the 
form~\cite{Serre_53,Su_Gardner_1969,Green_74,Green_76}: 
\begin{equation}
 \label{governing_equations}
\begin{aligned}
& h_t + (hu)_x = 0,\\
& (hu)_t + (hu^2 + p)_x = 0,\\
& \left(h e \right)_t + \left(hue+pu\right)_x = 0,
\end{aligned}
\end{equation}
with 
\begin{equation}
 \label{pressure_and_energy}
p = \frac{gh^2}{2} + \frac{1}{3}h^2\frac{D^2h}{Dt^2}, \quad 
e = \frac{u^2}{2} + \frac{gh}{2} + \frac{1}{6}\left(\frac{Dh}{Dt}\right)^2, \quad 	
\frac{D}{Dt} = \pd{}{t} + u\pd{}{x}.
\end{equation}
Here  $h$ is the
fluid depth, $u$ is the averaged over the fluid depth velocity, 
$p$ is the integrated over the fluid depth pressure. 
If $L_0$ is a characteristic wave length, 
and $H_0$ is the characteristic water depth, we define 
the dimensionless small parameter $\beta= H_0^2 /L_0^2$. The
SGN equations are obtained by depth-averaging the Euler system and
keeping in the resulting set of equations only first order terms in $\beta$ 
without making any assumptions on the amplitude of the waves.  
The third  equation in~\eqref{governing_equations} (the energy equation) is a consequence of 
the mass and momentum equations (the first two equations). The fourth conservation law 
(generalized Bernoulli conservation law) can also be written here
(cf.~\cite{Gavrilyuk_Teshukov_2001,Gavrilyuk_2015}).

Mathematical justification of this model and some related systems can be found 
in~\cite{Makarenko_1986,Li_2006,Saut_2012,LannesBOOK_2013,Duchene_2018}. 
A variational formulation of the SGN equations is given in~\cite{Salmon,Li_2001,Gavrilyuk_Teshukov_2001}. 
The linear stability of solitary waves of small amplitude to the SGN equations was established in~\cite{Li_2001}.
Also, it has been mentioned there that numerically, the solitary waves are  stable for any wave amplitude. 
Recent years have seen increased activity in both the study of qualitative properties of the solutions to 
the SGN system  and in the development of numerical discretization 
techniques~\cite{Gavrilyuk_Teshukov_2004,LGLX_2014,Metayer10,Favrie_Gavrilyuk_2017,Duran_2017,Marche_2018,Gavrilyuk_2018}.

In~\cite{El06} the Riemann problem for the SGN equations was examined. 
Earlier, the Riemann problem was mainly studied for integrable systems in~\cite{Gurevich_Pitaevskii_1974} 
for the KdV equation, and in~\cite{Gurevich_Krylov_1987,El_Geogjaev_Gurevich_Krylov_1995} for the NLS equation. 
Recently, this problem has received much attention for non-integrable systems of equations mainly 
because of the dispersive shocks commonly present in physics~\cite{El_2016}.  In~\cite{El06}  the wave number,  amplitude, average fluid depth 
and  average velocity have been chosen as primary variables to study the dispersive shocks of 
the SGN equations.  This choice is quite natural, because, for example, the leading edge of the 
dispersive shock corresponds to the limit of small wave numbers, while the trailing edge is 
the limit of small amplitudes. So, the asymptotic study in the limit of small wave numbers or 
small amplitudes is important to predict the solution behaviour. 

To capture better the case of moderate and large amplitude waves 
one can try to use  other variables. Even if {\it a priori} they may  be not necessarily physically tractable,
they could be useful to parametrize globally the generic solution to the modulation equations. 
In particular, it could help to determine the regions of modulational stability and instability 
for waves of arbitrary  amplitude. 

The structure of the article is as follows. The system of four modulation equations is 
derived in Sections~$2$,~$3$,~$4$. In Section~$5$ the averaged quantities are expressed as functions of 
the roots of the third order polynomial  determining  the fluid depth behaviour, and the phase velocity. 
The non-conservative form of the modulation equations and their hyperbolicity analysis are  given in 
Sections~$6$,~$7$. Numerical tests showing the wavetrain stability for the full SGN equations are 
presented in Section~$8$. Technical details are described in Appendix.

\section{Averaging of the conservation laws of the  SGN equations}
 A formal derivation of the modulation equations to the SGN equations 
(even in a more general formulation which contained, in particular, equations of bubbly fluids) 
can be found in~\cite{Gavrilyuk_1994,Gavrilyuk_2018}. 
However, the analysis of the hyperbolicity for such a general formulation was not performed there.
Here we will concentrate on SGN equations and will use the approach based on the averaging of 
conservation laws. 
	
Suppose that the unknowns $h$, $u$ (and also $p$ and $e$ which are functions of these variables 
and their derivatives) depend on the rapid travelling coordinate $\xi = x-Dt$ and 
slow variables $X=\varepsilon x$, $T=\varepsilon t $ (see Appendix~\ref{appendix_multiscale} for the details).
Here $D$ is the travelling wave velocity. Let us introduce the following 
$\varepsilon$--expansion ansatz for $h$, $u$, $p$ and $e$:
\begin{equation*}
 \begin{aligned}
   h(\xi,X,T) & = h_0(\xi,X,T)+\varepsilon h_1(\xi,X,T)+O(\varepsilon^2),\\
   u(\xi,X,T) & = u_0(\xi,X,T)+\varepsilon u_1(\xi,X,T)+O(\varepsilon^2),\\
   p(\xi,X,T) & = p_0(\xi,X,T)+\varepsilon p_1(\xi,X,T)+O(\varepsilon^2),\\
   e(\xi,X,T) & = e_0(\xi,X,T)+\varepsilon e_1(\xi,X,T)+O(\varepsilon^2).\\
\end{aligned}
\end{equation*}
Here all the terms are supposed to be $L$-periodic with respect to $\xi$, 
where $L$ is also a slowly varying function of $X, T$. 
The substitution of these expansions into~\eqref{governing_equations} yields:
\begin{equation*}
 \begin{aligned}
   & -Dh_{0\xi} + (h_0u_0)_\xi+\varepsilon\big(h_{0T} + (h_0u_0)_X\big) = 
          -\varepsilon\big[-Dh_{1\xi}+(h_0u_1+h_1u_0)_\xi\big] + O(\varepsilon^2),\\
   & -D(h_0u_0)_\xi + (h_0u_0^2 + p_0)_\xi+
          \varepsilon\big( (h_0u_0)_T + (h_0u_0^2 + p_0)_X\big) = \\
   & \qquad   -\varepsilon\big[-D(h_0u_1+h_1u_0)_\xi+(2h_0u_0u_1+h_1u_0^2 +p_1)_\xi\big] + 
                 O(\varepsilon^2),\\
   & -D(h_0 e_0 )_\xi + \left(h_0u_0e_0+p_0u_0\right)_\xi+\varepsilon\big((h_0 e_0 )_T + 
        \left(h_0u_0e_0+p_0u_0\right)_X\big)= \\
   & \qquad  \varepsilon \big[-D(h_1e_0+h_0e_1)_\xi+
      (h_1u_0e_0+h_0u_1e_0+h_0u_0e_1+p_0u_1+p_1u_0)_\xi\big] + O(\varepsilon^2).
\end{aligned}
\end{equation*}
Here we took into account the following transformations of the partial derivatives 
with respect to time and space:
\begin{equation}
   \label{the_derivatives}
	\pd{}{t} = -D \pd{}{\xi}+\varepsilon \pd{}{T}, \qquad 
        \pd{}{x} =  \pd{}{\xi}+\varepsilon \pd{}{X}.
\end{equation}
In particular, the material derivative $\frac{D}{Dt}$ reads:
\begin{equation*}
 \frac{D}{Dt} = \pd{}{t} + u\pd{}{x} = \big(u-D\big)\pd{}{\xi} + \varepsilon \Big(\pd{}{T}+u\pd{}{X}\Big).
\end{equation*}
Let us consider only the zero and first order approximation with respect to $\varepsilon$. 
The zero-order system reads:
\begin{equation}
 \label{zero_order_part}
 \begin{aligned}
    & -Dh_{0\xi} + (h_0u_0)_\xi=0,\\
    & -D(h_0u_0)_\xi + (h_0u_0^2 + p_0)_\xi = 0,\\
    & -D(h_0 e_0 )_\xi + \left(h_0u_0e_0+p_0u_0\right)_\xi=0.
 \end{aligned}
\end{equation}
The first-order system reads:
\begin{equation*}
 \begin{aligned}
  & h_{0T} + (h_0u_0)_X = -\big[-Dh_{1\xi}+(h_0u_1+h_1u_0)_\xi\big],\\
  & (h_0u_0)_T + (h_0u_0^2 + p_0)_X = 
    -\big[-D(h_0u_1+h_1u_0)_\xi+(2h_0u_0u_1+h_1u_0^2 +p_1)_\xi\big],\\
  & (h_0 e_0 )_T + \left(h_0u_0e_0+p_0u_0\right)_X=  
    \big[-D(h_1e_0+h_0e_1)_\xi+
    (h_1u_0e_0+h_0u_1e_0+h_0u_0e_1+p_0u_1+p_1u_0)_\xi\big].
\end{aligned}
\end{equation*}
Since all the functions $h_i$, $u_i$, $e_i$, and $p_i $,
$i=0,1$, are $L$-periodic with respect to $\xi$,  
after averaging the first order equations over the period $L$ 
one gets the following system:
\begin{equation}
 \label{first_order_part}
  \begin{aligned}
   & (\overline{h_0})_T + (\overline{h_0u_0})_X =0, \\
   & (\overline{h_0u_0})_T + (\overline{h_0u_0^2} + 
      \overline{p_0})_X =0,\\
   & (\overline{h_0 e_0} )_T + 
      \left(\overline{h_0u_0e_0}+\overline{p_0u_0}\right)_X=0.
  \end{aligned}
\end{equation}
Notice that here we used the fact that the averaging procedure 
and the derivation with respect to slow variables commute 
(cf.~\cite{Whitham_1974,Bhatnagar_1979,Kamchatnov_2000}).
	
\section{Stationary periodic solution}
\label{stp}
We will now show that the equations of zero order
approximation~\eqref{zero_order_part} admit periodic solutions.
We rewrite~\eqref{zero_order_part} in the form:
\begin{equation*}
 \begin{aligned}
	& -Dh' + (hu)'=0,\\
	& -D(hu)' + (hu^2 + p)' = 0,\\
	& -D(he)' + \left(hue+pu\right)'=0.
 \end{aligned}
\end{equation*}
Here and further the zero index is omitted and primes 
stand for $\pd{}{\xi}$. The first equation reads:
\begin{equation*}
  -Dh' + (hu)' = 0.
\end{equation*}
The integration gives:
\begin{equation}
  \label{full_mass}
  h(u-D) = m = const.
\end{equation}
 When we write here and further, $ m = const$, for example, we mean that $m$ does not depend on rapid variable $\xi$ : it is  a function of just two variables $T$ and $X$. The second equation can be integrated as :
\begin{equation}
  \label{full_momentum}
  p = i-\frac{m^2}{h}, \quad i=const.
\end{equation}
The second-derivative term in~\eqref{pressure_and_energy} 
can be transformed. Keeping only the zero powers of 
$\varepsilon$, we rewrite the second derivative of $h$ as:
\begin{equation*}
  \frac{D^2h}{Dt^2}=(u-D)\big((u-D)h'\big)'=
   \frac{(u-D)h}{h}\left(\frac{(u-D)h}{h}h'\right)'=
   \frac{m}{h}\left(\frac{mh'}{h}\right)=
    \frac{m^2}{h}\left(\frac{h'}{h}\right)'.
\end{equation*}
Thus, we have
\begin{equation*}
  \frac{D^2h}{Dt^2}=\frac{m^2}{h}\left(\frac{h'}{h}\right)'.
\end{equation*}
Hence, the pressure expression in~\eqref{pressure_and_energy} reads:
\begin{equation*}
  p = \frac{gh^2}{2} + \frac{1}{3} m^2 h \left(\frac{h'}{h}\right)'.
\end{equation*}
Replacing the pressure expression into~\eqref{full_momentum} 
one obtains:
\[
 \frac{m^2}{h} + \frac{gh^2}{2} + 
   \frac{1}{3}m^2h\left(\frac{h'}{h}\right)' = i.
\]
Multiplying both side of the equation by $h^{'}/m^{2} h^{2}$,
we have
\[
   \frac{h'}{h^3}+\frac{gh'}{2m^2} +
   \frac{1}{3} \frac{h'}{h}\left(\frac{h'}{h}\right)' = 
    \frac{ih'}{m^2h^2},
\]
which can be rewritten in the form:
\[
 \frac{1}{6}\left[\left(\frac{h'}{h}\right)^2\right]' + 
   \frac{h'}{h^3} + \frac{gh'}{2m^2} = \frac{ih'}{m^2h^2}.
\]
Integrating the equation once leads to 
\[
 \frac{1}{6}\left(\frac{h'}{h}\right)^2 - \frac{1}{2h^2} + 
  \frac{gh}{2m^2} = - \frac{i}{m^2h} + \epsilon,\quad \epsilon =const,
\]
Thus, the equation for $h$ is given  as:
\begin{equation}
 \label{main_oscillation_equation}
  \left ( h' \right )^2 = 
    3 - \frac{6i}{m^2}h + 6\epsilon h^2 - \frac{3g}{m^2}h^3.
\end{equation}	
Denote the polynomial in the right-hand side as $F_3(h)$ :
\begin{equation*}
 F_3(h) =  3 - \frac{6i}{m^2}h + 6\epsilon h^2 - 
    \frac{3g}{m^2}h^3,
\end{equation*}
or, equivalently:
\begin{equation*}
 F_3(h) = \frac{3g}{m^2}\left(\frac{m^2}{g} - 
    \frac{2i}{g}h + \frac{2\epsilon m^2}{g}h^2 - 
    h^3\right)=\frac{3g}{m^2} (h-h_0)(h-h_1)(h_2-h),
\end{equation*}
where $ h_0 < h_1 <  h_2 $ are the roots of $ F_3(h)$.
 Using Vieta's formulas one can write $ F_3(h) $ in the following way:
\begin{equation*}
	F_3(h) = \frac{3}{I_3}(I_3 - I_2h + I_1h^2 -h^3),
\end{equation*}
where
\begin{equation}
  \label{invariants}
\begin{aligned}
	& I_1 = h_0+h_1+h_2,\\
	& I_2 = h_0h_1 + h_1h_2 + h_0h_2,\\
	& I_3 = h_0h_1h_2.
\end{aligned}
\end{equation}
Identifying the coefficients of $F_3(h)$, one obtains :
\begin{equation}
 \label{vieta}
  \begin{aligned}
	 &I_1 = \frac{2 \epsilon m^2}{g},\\
	 &I_2 =\frac{2i}{g},\\
	 & I_3=\frac{m^2}{g}.
  \end{aligned}
\end{equation}
We are searching for the periodic solutions 
of~\eqref{main_oscillation_equation},  
oscillating between two real positive roots $h_1$ and $h_2$. 
Since they are real, the third root $h_0$ is real too. 
Moreover, the last formula in~\eqref{vieta} implies that 
$h_0$ is necessarily positive. 

The periodic solution that oscillates between $h_1$ and $h_2$ is given by the formula:
\begin{equation}
    h=h_1+(h_2-h_1) {\rm cn}^2( \alpha \, \xi ; k), \quad
     \alpha^2=\frac{3}{4}\frac{(h_2-h_0)}{h_0 h_1 h_2},
    \quad k^2=\frac{h_2-h_1}{ h_2-h_0}.
\label{explicit_solution}
\end{equation}
Here the Jacobi elliptic function $cn(u;k)$ is defined as:
\begin{equation*}
{\rm cn}(u;k)={\rm cos}(\varphi (u,k)),
\end{equation*}
where $\varphi (u,k)$ is obtained implicitly from the relation
\begin{equation*}
\int_0^\varphi \frac{d\theta}{\sqrt{1-k^2 \sin^2(\theta)}}=u.
\end{equation*}
The wavelength $L$ can explicitly be  given as:
\begin{equation*}
 \begin{aligned}
  L =  \int_{\xi_1}^{\xi_2}d\xi
    = 2 \int_{h_1}^{h_2} \frac{ dh}{\sqrt{F_3(h)}} 
    = 2 \sqrt{\frac{h_0h_1h_2}{3}} 
      \int_{h_1}^{h_2}\frac{dh}{\sqrt{P_3(h)}},
 \end{aligned}
\end{equation*}
where the interval $[\xi_1, \xi_2]$ has the length $L$ and
	\begin{equation}
	P_3(h)=(h-h_0)(h-h_1)(h_2-h).
	\label{polynomial_P}
	\end{equation}
The wavelength is thus completely defined by the roots $ h_0 $, $h_1 $ and $h_2 $. The averaging of any arbitrary function of $f(h)$ reads:
\begin{equation}
   \overline{f(h)} = \frac{1}{L} \int_{\xi_1}^{\xi_2}f(h)d\xi 
     = \frac{2}{L} \int_{h_1}^{h_2}\frac{f(h)dh}{\sqrt{F_3(h)}} 
     = \int_{h_1}^{h_2}\frac{f(h)dh}{\sqrt{P_3(h)}}\bigg/
       \int_{h_1}^{h_2}\frac{dh}{\sqrt{P_3(h)}}.
	\label{definition_averaging}
\end{equation} 
	
\section{Averaged equations}
Consider the first-order part of~\eqref{first_order_part}
(``zero" index is omitted):
\begin{equation}
 \label{averaged_governing_equations}
\begin{aligned}
 & (\overline{h})_T + (\overline{hu})_X =0,\\
 & (\overline{hu})_T + (\overline{hu^2} + \overline{p})_X =0,\\
 & (\overline{he} )_T + \left(\overline{hue}+\overline{pu}\right)_X=0.
\end{aligned}
\end{equation}
In the following, we will express all averaged quantities 
in~\eqref{averaged_governing_equations} in terms of 
four unknowns: $h_0$, $h_1$, $h_2$, and $D$.  

The flux in the first equation of \eqref{averaged_governing_equations} is:
\begin{equation*}
  \overline{hu} = \overline{h(u-D+D)} 
                = \overline{m+Dh} 
                = m + D\overline{h}=\overline{h}U, \quad 
   U=\displaystyle{\frac{\overline{hu}}{\overline{h}}}.
\end{equation*}
We introduced here the depth averaged velocity $U$. 
In terms of this velocity the mass equation can be rewritten in standard form:
\begin{equation}
 (\overline{h})_T + (\overline{h}U)_X =0.
 \label{standard_form}
\end{equation}
Since
\begin{equation*}
 \begin{aligned}
	\overline{hu^2}
	& = \overline{h(u-D+D)^2}\\
	& = \overline{h(u-D)^2} + 2\overline{h(u-D)}D + D^2\overline{h}\\
	& = \overline{\left(\frac{h^2(u-D)^2}{h}\right)} + 2\overline{h(u-D)}D + D^2\overline{h}\\
	& = m^2\overline{h^{-1}} + 2Dm + D^2\overline{h},
 \end{aligned}
\end{equation*}
and
\begin{equation*}
 \overline{p} = i - m^2 \overline{h^{-1}},
\end{equation*}
the flux in the second equation of~\eqref{averaged_governing_equations} is  :
\begin{equation*}
 \overline{hu^2+p} = i+ 2Dm + D^2\overline{h}=\overline{h}U^2+i-\frac{m^2}{\overline{h}}.
\end{equation*}
The last two terms represent a combination of the pressure (defined up to multiplicative constant which is the fluid density) first integrated over the water depth and 
then averaged over the wave period, 
and the corresponding quadratic velocity correlation. 
Together, the two terms form an ``effective pressure".  
One can prove that such an  effective pressure is always positive: 
\begin{equation*}
 i-\frac{m^2}{\overline{h}} = 
    \frac{g}{2}\left( I_2-2 I_3 \frac{\int\limits_{h_1}^{h_2}
    \frac{dh}{\sqrt{P_3(h)}}}{\int\limits_{h_1}^{h_2}\frac{hdh}{\sqrt{P_3(h)}}} \right)>0.
\end{equation*}
In the third equation,  we need to calculate  $ \overline{he}$, $\overline{hue}$, 
   and $\overline{pu}$. Let us remark that for the travelling wave solutions one has 
\begin{equation*}
  \left(\frac{Dh}{Dt}\right)^2 =\left((u-D)h'\right)^2=m^2\frac{h'^2}{h^2}.
\end{equation*} 
Also, it follows from~\eqref{main_oscillation_equation} that 
\begin{equation*}
 \overline{\left(\frac{h'^2}{h}\right)} = 
     -\frac{6i}{m^2}+ 3\overline{h^{-1}} + 6\epsilon \overline{h} - 
     \frac{3g}{m^2}\overline{h^2},
\end{equation*}
and
\begin{equation*}
  \overline{\left(\frac{h'}{h}\right)^2} = 
       6\epsilon + 3\overline{h^{-2}} - \frac{6i}{m^2}\overline{h^{-1}} - \frac{3g}{m^2}\overline{h}.
\end{equation*}
The averaged energy is:
\begin{equation*}
 \begin{aligned}
  \overline{e}
	& = \overline{\frac{u^2}{2} + \frac{gh}{2}+\frac{1}{6}\left(\frac{Dh}{Dt}\right)^2}\\
	& = \overline{\frac{u^2}{2} + \frac{gh}{2}+\frac{1}{6}\big((u-D)h'\big)^2}\\
	& = \frac{1}{2}\overline{\big((u-D)^2+2D(u-D)+D^2\big)} + 
            \frac{g}{2}\overline{h}+\frac{1}{6}\overline{\left(h^2(u-D)^2 \frac{h'^2}{h^2}\right)}\\
	& = \frac{1}{2} \overline{\left(\frac{h^2(u-D)^2}{h^2}\right)} + 
            D\overline{\left( \frac{h(u-D)}{h}\right)}+\frac{1}{2}D^2+\frac{g}{2}\overline{h} + 
            \frac{1}{6}m^2\overline{\left(\frac{h'^2}{h^2}\right)} \\
	& = \frac{1}{2}m^2\overline{h^{-2}} + Dm\overline{h^{-1}}+\frac{1}{2}D^2 + 
            \frac{g}{2}\overline{h} + \frac{m^2}{6}\overline{\left(\frac{h'}{h}\right)^2}.
 \end{aligned}
\end{equation*}
Thus, the averaged energy reads:
\begin{equation*}
  \overline{e} = \frac{1}{2}D^2+m^2\epsilon + m^2\overline{h^{-2}}+(Dm-i)\overline{h^{-1}}.
\end{equation*}
It can be also expressed as a function of $U$ (instead of $D$) and $h_0$, $h_1$, $h_2$. 
The volume average energy is:
\begin{equation*}
 \begin{aligned}
   \overline{he}
    & = \overline{\frac{hu^2}{2} + \frac{gh^2}{2} + \frac{h}{6}\left(\frac{Dh}{Dt}\right)^2}\\
    & = \frac{1}{2} \left(2Dm + m^2\overline{h^{-1}} + D^2\overline{h}\right)+\overline{\frac{gh^2}{2} + 
        \frac{h}{6}\big((u-D)h'\big)^2}\\
    & = \frac{1}{2} \left(2Dm + m^2\overline{h^{-1}} + D^2\overline{h}\right)+
        \frac{g}{2}\overline{h^2} + \frac{1}{6}\overline{\left(\frac{h^2(u-D)^2h'^2}{h}\right)}\\
    & = \frac{1}{2} \left(2Dm + m^2\overline{h^{-1}} + D^2\overline{h}\right)+
        \frac{g}{2}\overline{h^2} + \frac{m^2}{6} \overline{\left(\frac{h'^2}{h}\right)}.
 \end{aligned}
\end{equation*}
Thus,
\begin{equation*}
 \overline{he} = Dm-i+m^2\overline{h^{-1}}+\left(\frac{1}{2}D^2+m^2\epsilon\right)\overline{h},
\end{equation*}
and
\begin{equation*}
 \begin{aligned}
 \overline{hue} & = \overline{h(u-D+D)e} = \overline{me + Dhe} = m\overline{e} + D\overline{he} \\
                & = \frac{3}{2}D^2m-iD+m^3\epsilon+m^3\overline{h^{-2}}+
                    (2Dm^2-mi)\overline{h^{-1}}+\left(\frac{1}{2}D^3+m^2D\epsilon\right)\overline{h}.
 \end{aligned}
\end{equation*}
Alos, one has :
\begin{equation*}
 \begin{aligned}
  \overline{pu} 
	& = \overline{iu - m^2uh^{-1}}\\
	& = \overline{i \frac{h}{h} (u-D+D) - m^2 \frac{h}{h^2}(u-D+D)}\\
	& = \overline{\frac{i}{h}h(u-D)} + iD - \overline{\frac{m^2}{h^2}h(u-D)} - m^2 D \overline{h^{-1}}\\
	& = iD - m^3\overline{h^{-2}} - (Dm^2-mi)\overline{h^{-1}}.
\end{aligned}
\end{equation*}
Now we are able to write the modulation equations because all quantities we need are given explicitly:
\begin{equation}
 \label{macro_quantities_thermo}
 \begin{aligned}
   \overline{hu} & = \overline{h}D + m,\\
   \overline{hu^2+p} & = \overline{h}D^2 + 2mD +i,\\
   \overline{he} & = \frac{1}{2}\overline{h}D^2 +mD-i+ m^2\epsilon\overline{h}+m^2\overline{h^{-1}},\\
   \overline{hue+pu} & = \frac{1}{2}\overline{h}D^3+m^2\epsilon\overline{h}D+ 
                      \frac{3}{2}mD^2+m^3\epsilon+m^2\overline{h^{-1}}D.\\
 \end{aligned}
\end{equation}
The last step would be to replace the integration constants 
$m$, $i$ and $\epsilon$ by their expressions in terms of invariants $I_i$, $ i=1,2,3 $, 
using~\eqref{vieta}:
\begin{equation}
 \label{constants_invariants}
 \begin{aligned}
   m^2      & = gI_3 \; (m=\sgn{m}\sqrt{gI_3}),\\
   i        & = \frac{1}{2}gI_2,\\
   \epsilon & =\frac{1}{2}\frac{I_1}{I_3}.
 \end{aligned}
\end{equation}
The negative (positive) sign of $m$ corresponds to the right (left) facing periodic waves.  
We introduce the synthesis of both notations in order to obtain the simplest form of the equations. 
Basically, we will describe everything in terms of $m$, $i$ and $I_1$.  
Using~\eqref{constants_invariants}, one can eliminate the dependence on $\epsilon$ 
in~\eqref{macro_quantities_thermo}:	
\begin{equation}
 \label{macro_quantities_thermo_new}
 \begin{aligned}
   \overline{hu}     & = \overline{h}D + m,\\
   \overline{hu^2+p} & = \overline{h}D^2 + 2mD +i,\\
   \overline{he}     & = \frac{1}{2}\overline{h}D^2 +mD-i+ 
                         \frac{1}{2}g I_1\overline{h}+m^2\overline{h^{-1}},\\
   \overline{hue+pu} & = \frac{1}{2}\overline{h}D^3+\frac{1}{2}g I_1\overline{h}D+ 
                         \frac{3}{2}mD^2+\frac{1}{2}g I_1m+m^2\overline{h^{-1}}D.\\
 \end{aligned}
\end{equation}
Complemented by equation~\eqref{wave_length_equation} for the 
wavelength (see Appendix~\ref{appendix_multiscale}),  equations~\eqref{first_order_part} will finally be written as:
\begin{equation}
 \label{modulation_equations_pre}
 \begin{aligned}
  & L_T - LD_X + DL_X = 0,\\
  & \overline{h}_T+\left(m+\overline{h}D\right)_X = 0,\\
  & \left(m+\overline{h}D\right)_T+\left(\overline{h}D^2 +\frac{1}{2}gI_2+2mD\right)_X=0,\\
  & \left(\frac{1}{2}\overline{h}D^2+\frac{1}{2}gI_1\overline{h}-
     \frac{1}{2}gI_2+gI_3\overline{h^{-1}}+mD\right)_T+\\
  & \qquad \left(\frac{1}{2}\overline{h}D^3+\frac{1}{2}gI_1\overline{h}D+gI_3\overline{h^{-1}}D+
       \frac{3}{2}m D^2+\frac{1}{2}m gI_1\right)_X=0.
 \end{aligned}
\end{equation}
We need now to rewrite~\eqref{modulation_equations_pre} in quasilinear form in variables 
$D$, $h_0$, $h_1$, and $h_2$. 

\section{Expressions for the main averaged variables}
The expressions of  $\overline{h}$, $\overline{h^{-1}}$ and $L$ in terms of $h_0$, $h_1$,
and $h_2$ are (for proof see Appendix~\ref{appendix_definitions}):
\begin{equation*}
  \overline{h} = h_0 + (h_2-h_0)\frac{E(k)}{K(k)}, \quad \overline{h^{-1}} = \frac{\Pi(n,k)}{h_2K(k)}, \quad 
	L = 4 \sqrt{\frac{h_0h_1h_2}{3}} \frac{K(k)}{\sqrt{h_2-h_0}}.
\end{equation*}
Then, one can write  the following differentials : 
\begin{equation*}
 \begin{aligned}
  d\overline{h}      & = \Phi^0 dh_0 + \Phi^1 dh_1 + \Phi^2 dh_2,\\
  d\overline{h^{-1}} & = \Psi^0 dh_0 + \Psi^1 dh_1 + \Psi^2 dh_2,\\
  dL                 & = \Lambda^0 dh_0 + \Lambda^1 dh_1 + \Lambda^2 dh_2,\\
  dI_1               & = dh_0 + dh_1 + dh_2,\\
  dI_2               & = (h_{1}+h_{2})dh_0 + (h_{0}+h_{2})dh_1 + (h_{0}+h_{1})dh_2,\\
  dI_3               & = h_{1}h_{2}dh_0 + h_{0}h_{2}dh_1 + h_{0}h_{1}dh_2,\\
  dm                 & = \frac{m}{2}\left(\frac{dh_0}{h_0}+\frac{dh_1}{h_1}+\frac{dh_2}{h_2}\right).
 \end{aligned}
\end{equation*}
Here $\Phi^i$, $\Psi^i$ and $\Lambda^i$ ($ i = 0,1,2 $) read 
(for proof see  Appendix~\ref{appendix_differentials}):
\begin{equation*}
 \begin{aligned}[c]
  \Phi^0 & = \frac{1}{2} - \frac{h_2-h_0}{2(h_1-h_0)} \frac{E^2(k)}{K^2(k)},\\
  \Phi^1 & = \frac{h_2-h_0}{2(h_2-h_1)} - \frac{h_2-h_0}{h_2-h_1}\frac{E(k)}{K(k)} + 
             \frac{(h_2-h_0)^2}{2(h_2-h_1)(h_1-h_0)}\frac{E^2(k)}{K^2(k)},\\
  \Phi^2 & = -\frac{h_1-h_0}{2(h_2-h_1)}+\frac{h_2-h_0}{h_2-h_1}\frac{E(k)}{K(k)} - 
              \frac{h_2-h_0}{2(h_2-h_1)}\frac{E^2(k)}{K^2(k)},\\
  \Psi^0 & = \frac{1}{2h_0(h_1-h_0)}\frac{E(k)}{K(k)}-\frac{1}{2h_0h_2}\frac{\Pi(n,k)}{K(k)} - 
             \frac{1}{2h_2(h_1-h_0)}\frac{\Pi(n,k)E(k)}{K^2(k)},\\
  \Psi^1 & = \frac{1}{2h_1(h_2-h_1)} - \frac{h_2-h_0}{2h_1(h_2-h_1)(h_1-h_0)}\frac{E(k)}{K(k)} - 
             \frac{1}{2h_1(h_2-h_1)}\frac{\Pi(n,k)}{K(k)}+ \\
         & \qquad  \frac{h_2-h_0}{2h_2(h_2-h_1)(h_1-h_0)}\frac{\Pi(n,k)E(k)}{K^2(k)},\\
  \Psi^2 & = -\frac{1}{2h_2(h_2-h_1)}+\frac{1}{2h_2(h_2-h_1)}\frac{E(k)}{K(k)} + 
             \frac{h_1}{2h_2^2(h_2-h_1)}\frac{\Pi(n,k)}{K(k)} - \frac{1}{2h_2(h_2-h_1)}
             \frac{\Pi(n,k)E(k)}{K^2(k)},\\
\Lambda^0 & = \frac{2}{\sqrt{3}}\left(\frac{\sqrt{h_0h_1h_2}}{(h_1-h_0)\sqrt{h_2-h_0}}E(k)+\frac{h_1h_2}{\sqrt{h_2-h_0}\sqrt{h_0h_1h_2}}K(k)\right),\\
\Lambda^1&= \frac{2}{\sqrt{3}}\left(-\frac{\sqrt{h_2-h_0}\sqrt{h_0h_1h_2}}{(h_2-h_1)(h_1-h_0)}E(k)+\frac{h_0h_2^2}{(h_2-h_1)\sqrt{h_2-h_0}\sqrt{h_0h_1h_2}}K(k)\right),\\
\Lambda^2&=\frac{2}{\sqrt{3}}\left(\frac{\sqrt{h_0h_1h_2}}{(h_2-h_1)\sqrt{h_2-h_0}}E(k)-\frac{h_0h_1^2}{(h_2-h_1)\sqrt{h_2-h_0}\sqrt{h_0h_1h_2}}K(k)\right).
\end{aligned}
\end{equation*}
The formulas for $\Phi^k$, $\Psi^k$ and $\Lambda^k$, $k=0,1,2$ were verified  
by hand calculations and with Wolfram Mathematica. One must pay attention to the fact that 
the complete elliptic integrals we use depend on elliptic modulus $k$, while Wolfram Mathematica 
uses the definition from Abramovitz and Stegun~\cite{AS_1964} where the complete elliptic integrals 
depend on parameter $m=k^2$ (do not confound the notations $m$ with $m$ coming from 
the mass conservation equation).

\section{Nonconservative modulation equations}

Complemented by~\eqref{consistency_equation_slow} (or, equivalently, by  equation~\eqref{wave_length_equation}) 
(see Appendix~\ref{appendix_multiscale}) the modulation equations~\eqref{modulation_equations_pre} 
can be written in the following developed form: 
\begin{equation}
  \label{final_equations}
 \begin{aligned}
   & \Lambda^0 h_{0T} + \Lambda^1 h_{1T} + \Lambda^2 h_{2T} - LD_X + D \Lambda^0 h_{0X} + 
       D \Lambda^1 h_{1X} + D \Lambda^2 h_{2X} = 0,\\
   & \Phi^0 h_{0T} + \Phi^1 h_{1T} + \Phi^2 h_{2T} +\overline{h}D_X + \Big(D\Phi^0 + 
      \frac{m}{2h_0}\Big)h_{0X}+ \Big(D\Phi^1 + \frac{m}{2h_1}\Big)h_{1X} + \Big(D\Phi^2 + 
         \frac{m}{2h_2}\Big)h_{2X} = 0,\\
   & \overline{h}D_T + \Big(D\Phi^0 + \frac{m}{2h_0}\Big)h_{0T}+ \Big(D\Phi^1 +
          \frac{m}{2h_1}\Big)h_{1T} + \Big(D\Phi^2 + \frac{m}{2h_2}\Big)h_{2T} +\\
   & \qquad  \left(2\overline{h}D+2m\right)D_X +  \Big(D^2\Phi^0+\frac{1}{2}g(h_1+h_2)+
             \frac{m}{h_0}D\Big)h_{0X}+ \\
   & \qquad \Big(D^2\Phi^1+\frac{1}{2}g(h_0+h_2)+\frac{m}{h_1}D\Big)h_{1X} +
	\Big(D^2\Phi^2+\frac{1}{2}g(h_0+h_1)+\frac{m}{h_2}D\Big)h_{2X} = 0,\\
   & \left(\overline{h}D+m\right)D_T + \\
   & \qquad \left(\frac{1}{2}\left(D^2+gI_1\right)\Phi^0 + m^2\Psi^0 + 
            \frac{1}{2}g(\overline{h}-h_1-h_2)+gh_1h_2\overline{h^{-1}}+\frac{m}{2h_0}D\right)h_{0T} +\\
   & \qquad \left(\frac{1}{2}\left(D^2+gI_1\right)\Phi^1 + m^2\Psi^1 + 
           \frac{1}{2}g(\overline{h}-h_0-h_2)+gh_0h_2\overline{h^{-1}}+\frac{m}{2h_1}D\right)h_{1T} +\\
   & \qquad \left(\frac{1}{2}\left(D^2+gI_1\right)\Phi^2 + m^2\Psi^2 + 
        \frac{1}{2}g(\overline{h}-h_0-h_1)+gh_0h_1\overline{h^{-1}}+\frac{m}{2h_2}D\right)h_{2T} + \\
   & \qquad \left(\frac{3}{2}\overline{h}D^2 + \frac{1}{2}gI_1\overline{h} + 
       m^2\overline{h^{-1}} + 3mD\right)D_X+\\
   & \qquad \left(\frac{1}{2}(D^2+gI_1)D\Phi^0 + m^2D\Psi^0 + \frac{1}{2}g\overline{h}D + 
       gh_1h_2\overline{h^{-1}}D+\frac{3}{4}\frac{m}{h_0}D^2+\frac{1}{4}g\frac{mI_1}{h_0} + 
       \frac{1}{2}gm\right) h_{0X} +\\
   & \qquad \left(\frac{1}{2}(D^2+gI_1)D\Phi^1 + m^2D\Psi^1 + \frac{1}{2}g\overline{h}D +
        gh_0h_2\overline{h^{-1}}D+\frac{3}{4}\frac{m}{h_1}D^2+\frac{1}{4}g\frac{mI_1}{h_1} + 
             \frac{1}{2}gm\right) h_{1X} +\\
   & \qquad \left(\frac{1}{2}(D^2+gI_1)D\Phi^2 + m^2D\Psi^2 + 
     \frac{1}{2}g\overline{h}D +gh_0h_1\overline{h^{-1}}D+\frac{3}{4}\frac{m}{h_2}D^2+
       \frac{1}{4}g\frac{mI_1}{h_2} + \frac{1}{2}gm\right) h_{2X} = 0.
 \end{aligned}
\end{equation}
Or, in matrix form:
\begin{equation*}
	A\textbf{U}_T + B\textbf{U}_X = 0,
\end{equation*}
where
\begin{equation*}
\textbf{U} = 
	\begin{bmatrix}
	D\\
	h_0\\
	h_1\\
	h_2
	\end{bmatrix}, \qquad					
 A = \begin{bmatrix}
	0        & a_{12}  & a_{13}  & a_{14} \\
	0        & a_{22}  & a_{23}  & a_{24} \\
	a_{31} & a_{32} & a_{33}  & a_{34} \\
	a_{41} & a_{42} & a_{43}  & a_{44} \\
     \end{bmatrix}, \qquad
  B = \begin{bmatrix}
	b_{11} & b_{12}  & b_{13}  & b_{14} \\
	b_{21} & b_{22} & b_{23}  & b_{24} \\
	b_{31} & b_{32} & b_{33}  & b_{34} \\
	b_{41} & b_{42} & b_{43}  & b_{44} \\
       \end{bmatrix}.
\end{equation*}
The coefficients of $ A $ are given by:
\begin{equation*}
 \hspace{3ex}\begin{aligned}
  & a_{11} = 0, \quad a_{12} = \Lambda^0, \quad a_{13} = \Lambda^1, \quad a_{14} = \Lambda^2,\\
  & a_{21} = 0, \quad a_{22} = \Phi^0, \quad a_{23} = \Phi^1, \quad a_{24} = \Phi^2, \\
  & a_{31} = \overline{h}, \quad a_{32} = D\Phi^0 + \frac{m}{2h_0}, 
      \quad a_{33} = D\Phi^1 + \frac{m}{2h_1}, \quad a_{34} = D\Phi^2 + \frac{m}{2h_2},\\
  & a_{41} = \overline{h}D+m,\\
  & a_{42} = \frac{1}{2}\left(D^2+gI_1\right)\Phi^0 + m^2\Psi^0 + 
             \frac{1}{2}g(\overline{h}-h_1-h_2)+gh_1h_2\overline{h^{-1}}+\frac{m}{2h_0}D,\\
  & a_{43} = \frac{1}{2}\left(D^2+gI_1\right)\Phi^1 + m^2\Psi^1 + 
             \frac{1}{2}g(\overline{h}-h_0-h_2)+gh_0h_2\overline{h^{-1}}+\frac{m}{2h_1}D,\\
  & a_{44} = \frac{1}{2}\left(D^2+gI_1\right)\Phi^2 + m^2\Psi^2 + 
             \frac{1}{2}g(\overline{h}-h_0-h_1)+gh_0h_1\overline{h^{-1}}+\frac{m}{2h_2}D.
 \end{aligned}
\end{equation*}
The coefficients of $ B $ are given by:
\begin{equation*}
\hspace{3ex}\begin{aligned}
 & b_{11} = - L, \quad b_{12} = D \Lambda^0, \quad b_{13} =  D \Lambda^1, \quad b_{14} = D \Lambda^2,\\
 & b_{21} = \overline{h}, \quad b_{22} =  D\Phi^0 + \frac{m}{2h_0}, 
      \quad b_{23} = D\Phi^1 + \frac{m}{2h_1}, \quad b_{24} = D\Phi^2 + \frac{m}{2h_2},\\
 & b_{31} = 2\overline{h}D+2m,\\
 & b_{32} = D^2\Phi^0+\frac{1}{2}g(h_1+h_2)+\frac{m}{h_0}D,\\
 & b_{33} = D^2\Phi^1+\frac{1}{2}g(h_0+h_2)+\frac{m}{h_1}D,\\
 & b_{34} = D^2\Phi^2+\frac{1}{2}g(h_0+h_1)+\frac{m}{h_2}D,\\
 & b_{41} = \frac{3}{2}\overline{h}D^2 + \frac{1}{2}gI_1\overline{h} + m^2\overline{h^{-1}} + 3mD,\\
 & b_{42} = \frac{1}{2}(D^2+gI_1)D\Phi^0 + m^2D\Psi^0 + 
            \frac{1}{2}g\overline{h}D + gh_1h_2\overline{h^{-1}}D+\frac{3}{4}\frac{m}{h_0}D^2+
            \frac{1}{4}g\frac{mI_1}{h_0} + \frac{1}{2}gm,\\
 & b_{43} = \frac{1}{2}(D^2+gI_1)D\Phi^1 + m^2D\Psi^1 + \frac{1}{2}g\overline{h}D + 
            gh_0h_2\overline{h^{-1}}D+\frac{3}{4}\frac{m}{h_1}D^2+\frac{1}{4}g\frac{mI_1}{h_1} + 
            \frac{1}{2}gm,\\
 & b_{44} = \frac{1}{2}(D^2+gI_1)D\Phi^2 + m^2D\Psi^2 + 
            \frac{1}{2}g\overline{h}D + gh_0h_1\overline{h^{-1}}D+\frac{3}{4}\frac{m}{h_2}D^2+
            \frac{1}{4}g\frac{mI_1}{h_2} + \frac{1}{2}gm.
\end{aligned}
\end{equation*}
The characteristic eigenvalues are the roots of the fourth order polynomial 
\begin{equation}
{\rm det}\left(B-\lambda A\right)=0. 
\label{eigenvalues}
\end{equation}
To simplify the computations let us remark that the introduction of the depth averaged 
velocity $U = \overline{uh}/\overline{h}$
allows us to rewrite the mass conservation equation in standard form~\eqref{standard_form}. 
Then we can use the fact that the equations are Galilean invariant if  
the depth average velocity is used. So, to check the  hyperbolicity for any $U$ is equivalent 
to check the hyperbolicity for $U=0$. Since
\begin{equation*}
 U = \frac{m}{\overline{h}} + D, 
\end{equation*}
we will put into the coefficients of the matrices $A$ and $B$ the value of $D$  
corresponding to $U=0$:
\begin{equation*}
	D = -\frac{m}{\overline{h}}.
\end{equation*}
Let us also remark that for the hyperbolicity study, one can always take 
$h_0=1$ in the coefficients of the polynomial~\eqref{eigenvalues} (if $h_0 =\alpha >0$, 
all the eigenvalues will only be multiplied by $\sqrt{\alpha}$). 
The corresponding symmetry relations are the consequences of the fact that the periodic solution is 
determined in terms of the third degree polynomial. 
	
\section{Hyperbolicity region}

\begin{figure}[!h]
\begin{center}
\includegraphics[width=10cm]{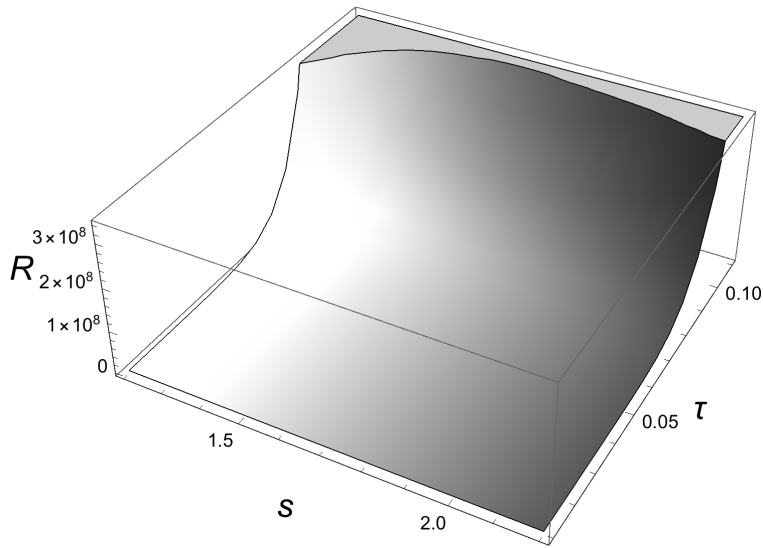}
\end{center}
\caption{The resultant $R$ of the polynomial~\eqref{eigenvalues} and 
 its derivative does not change sign in the region  
 $1<s<100$ and $0<\tau<100$ where all roots are real.  
 The resultant is plotted  here in a smaller region of $s$ and $\tau$. 
 Thus, the polynomial~\eqref{eigenvalues} has no multiple real roots. 
 It means that the system of modulation equations is 
 strictly hyperbolic. }
\label{Fig1}
\end{figure}

Since the roots of  polynomial~\eqref{polynomial_P} satisfy the inequality : $1=h_0<h_1<h_2$, 
one can parametrize $h_1$ and $h_2$ as : $h_1=s$, $h_2=s+\tau$, where $s>1$ and  $\tau>0$. 
Such a parametrization allows us  to use  a standard ``Cartesian" frame for the computation 
of the eigenvalues of  \eqref{eigenvalues}.  The eigenvalues thus  are  given  explicitly 
as functions of $s$ and $\tau$. We used  Wolfram  Mathematica for such a computation. 
The numerical results show that the eigenvalues are all real in a large region 
$\Omega=\{( s,\tau)\vert 1<s<100, \; 0<\tau<100\}$. 
First, for each pair $ (s,\tau ) $ from $\Omega$ we computed the numerical values of 
matrices $A$ and $B$. Then the corresponding  eigenvalues were computed as the roots 
of the fourth order polynomial~\eqref{eigenvalues}.  
Moreover, one can find that the resultant of the corresponding polynomial~\eqref{eigenvalues} 
and its derivative (denoted further as $R$) does not change sign  (see Fig.~\ref{Fig1}).  
Thus, the polynomial~\eqref{eigenvalues} has no multiple real roots.
Since all the roots are real and different,  the system of modulation equations for 
the SGN equations is thus {\it strictly  hyperbolic}.  The fact that  periodic waves of all lengths  
are modulationally stable corroborates the results of~\cite{Li_2001} where the spectral stability 
of solitary waves (the limit $s\rightarrow 1$) has been proven for small amplitudes 
and numerically confirmed for large amplitudes.  
	
\begin{figure}[!h]
\begin{center}
	\includegraphics[width=10cm]{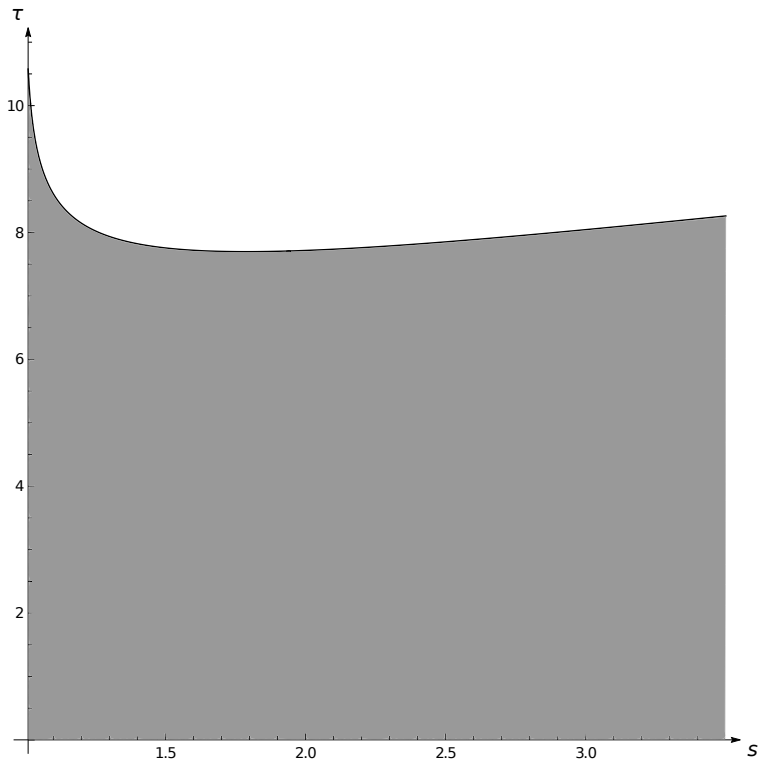}
\end{center}
\caption{The region  $s>1, \; \tau >0$  is divided by a smooth curve into two sub-regions, ``grey" and ``white". If $m<0$, in the grey sub-region one has three positive and one negative eigenvalues, while in the white sub-region one has two positive and two negative eigenvalues. If $m>0$, the signs of the roots will only change, since we have taken vanishing $U$.}
	\label{Fig2}
\end{figure}

The whole hyperbolicity region $s>1, \; \tau>0$ is divided by a smooth curve corresponding to 
$\lambda =0$ into two  sub-regions, ``grey" and ``white" (see Fig.~\ref{Fig2}). 
If $m<0$, in the grey sub-region one has three positive and one negative eigenvalues, 
while in the white sub-region one has two positive and two negative eigenvalues. 
If $m>0$, the signs of the roots will only change, since we have taken $U$ vanishing. 

\pagebreak

\section{Numerical results
         \label{sec:results}}

To study the modulational stability of periodic waves 
to SGN model numerically, we use the following set of parameters:
$h_{0} = 1\; m$, $h_{1}= 1.5\;m$, $h_{2} = 2\;m$ (i.e. 
$s = 1.5\; m$ and $\tau = 0.5\; m$),
and $g=10\; m/s^{2}$, as an example, for the solution of
the water height described by~\eqref{explicit_solution} over 
a single wave 
length $L$ which corresponds to $7.4163 \; m$ approximately.
To setup the problem, a wave train 
is formed initially that consists of
$N$ aforementioned single stationary wave solutions (see Section \ref{stp})
in a domain of size $L_{1} = N \times L$, where
the travelling wave speed of this wave train is taken be
$D = -m/\overline{h} \approx 3.1688\; m/s$ 
(this allows us to take vanishing the averaged over period the mass weighted velocity 
$U = \overline{hu}/\overline{h}$).
With that, we then introduce perturbations to the height of
the wave train $h(x)$ as 
\begin{subequations}
  \label{eq:sgn-mod-t0}
 \begin{align}
\tilde h(x) := h(x) \left (
         1 + a \cos \left ( \frac{2 \pi  x}{L_{1}}
                    \right )
                   \right ), 
 \end{align}
and define the perturbed velocity of the wave by
\begin{align}
 \tilde u(x): = \frac{m}{\tilde h(x)} + D,
 \end{align}
\end{subequations}
where $a$ is a small parameter.
For this problem,
periodic boundary conditions are assumed and used
on the left and right of
the interval $[0, L_{1}]$.

\begin{figure}[!h]
\begin{center}
\begin{tabular}{cc}
\begin{minipage}{2.6in}
\hskip 0.1in
\includegraphics[width=2.3in]{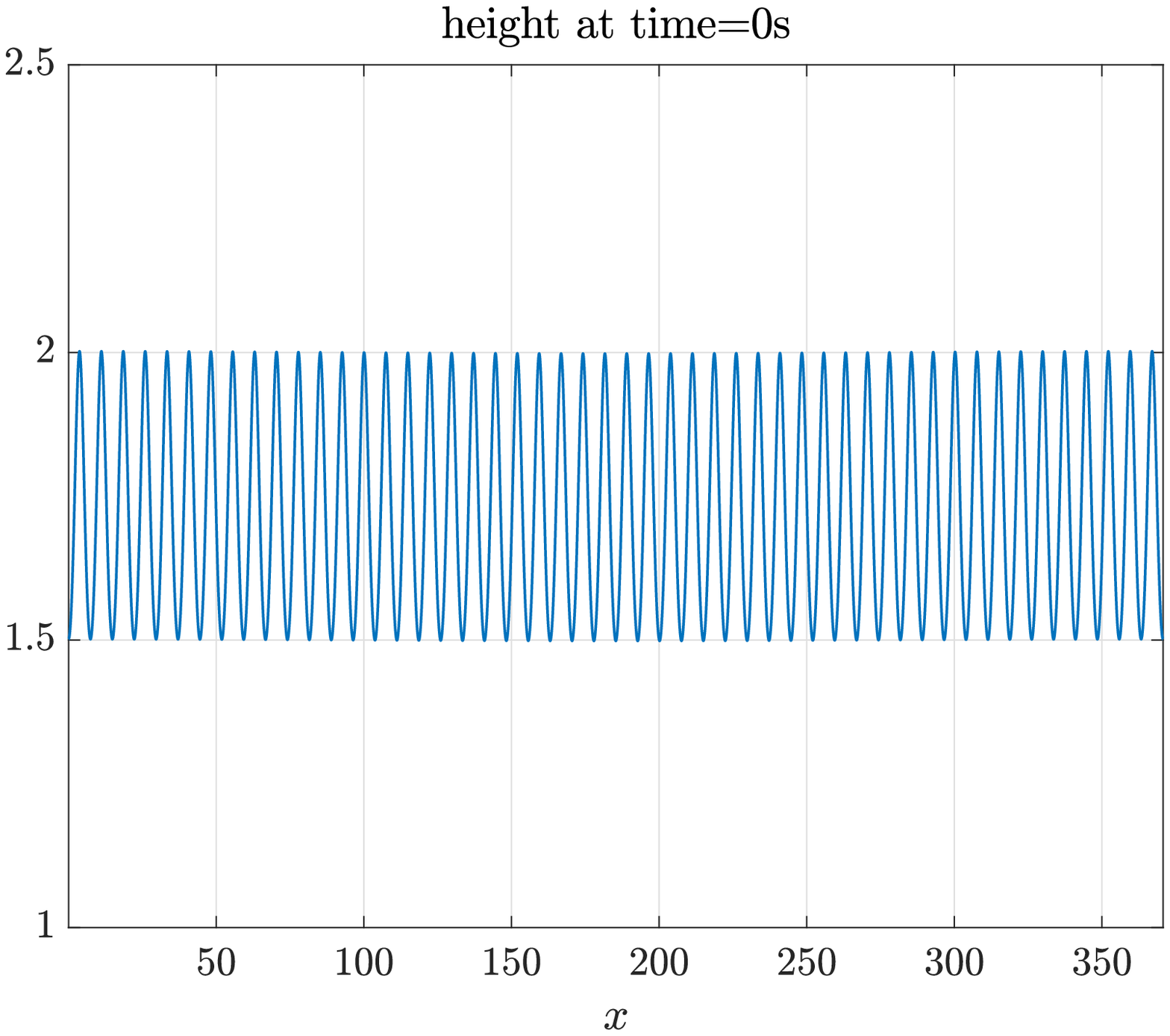}
\end{minipage}
&
\begin{minipage}{2.6in}
\includegraphics[width=2.5in]{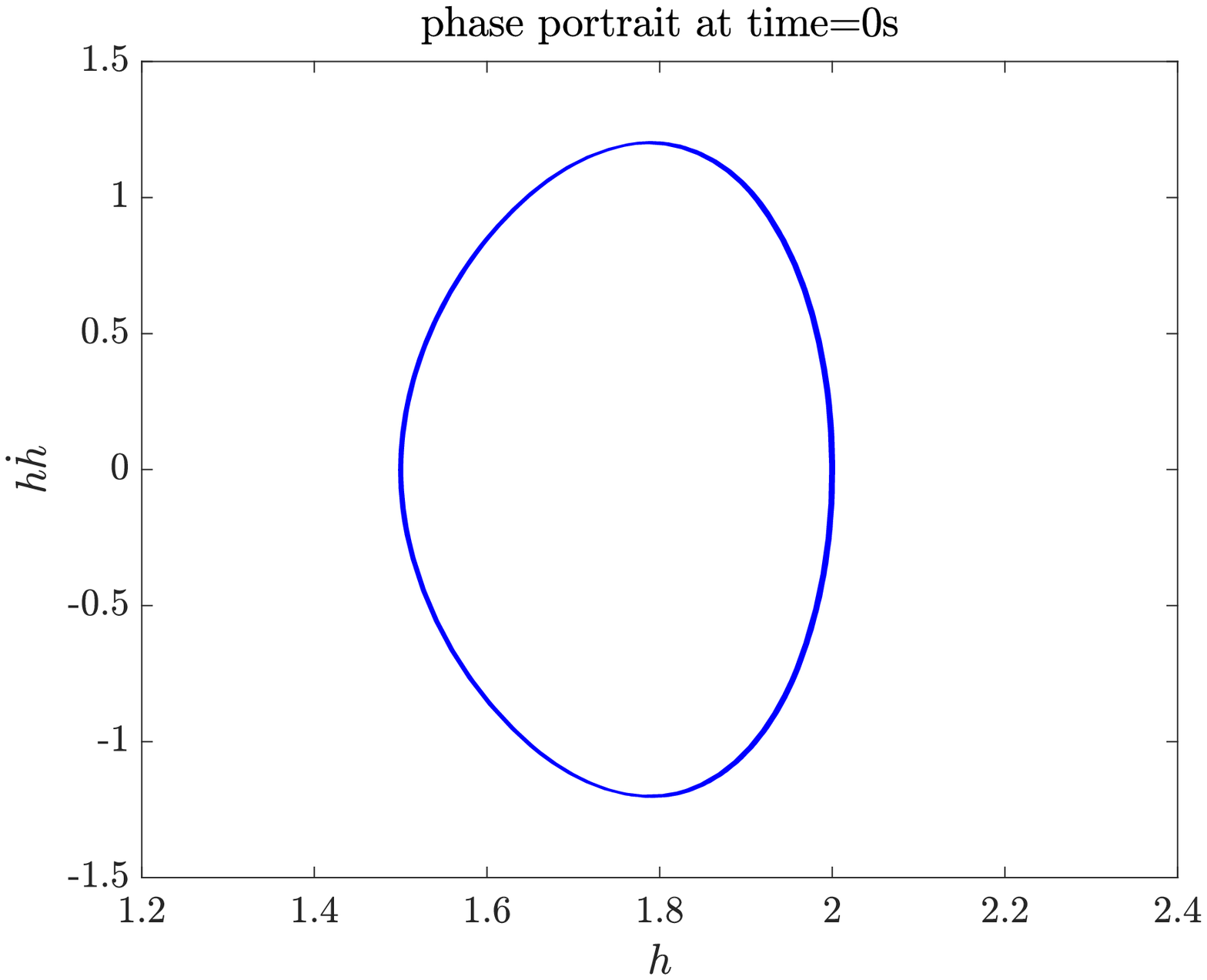}
\end{minipage}
\end{tabular}
\vskip 0.1in
\begin{tabular}{cc}
\begin{minipage}{2.6in}
\hskip 0.1in
\includegraphics[width=2.3in]{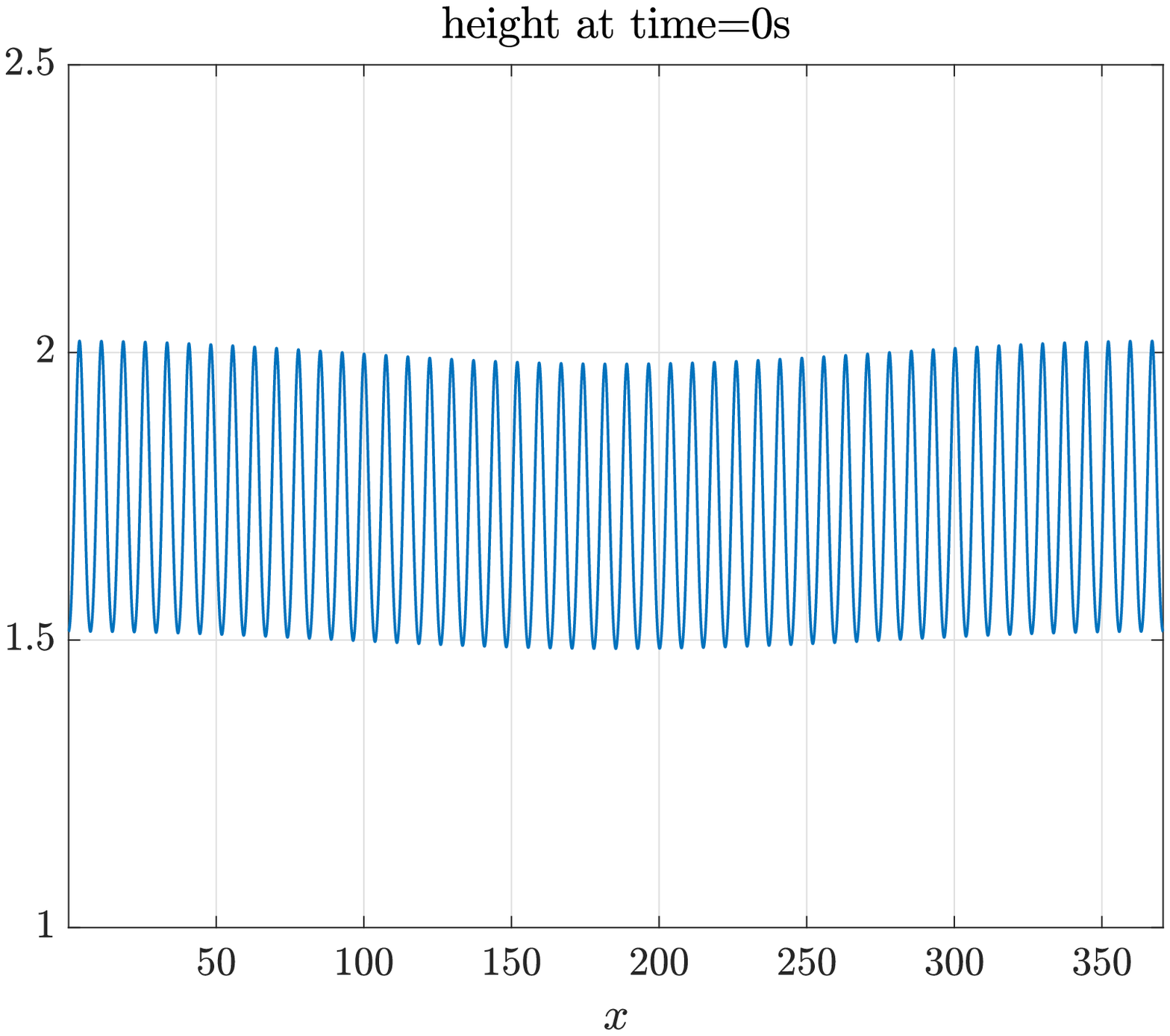}
\end{minipage}
&
\begin{minipage}{2.6in}
\includegraphics[width=2.5in]{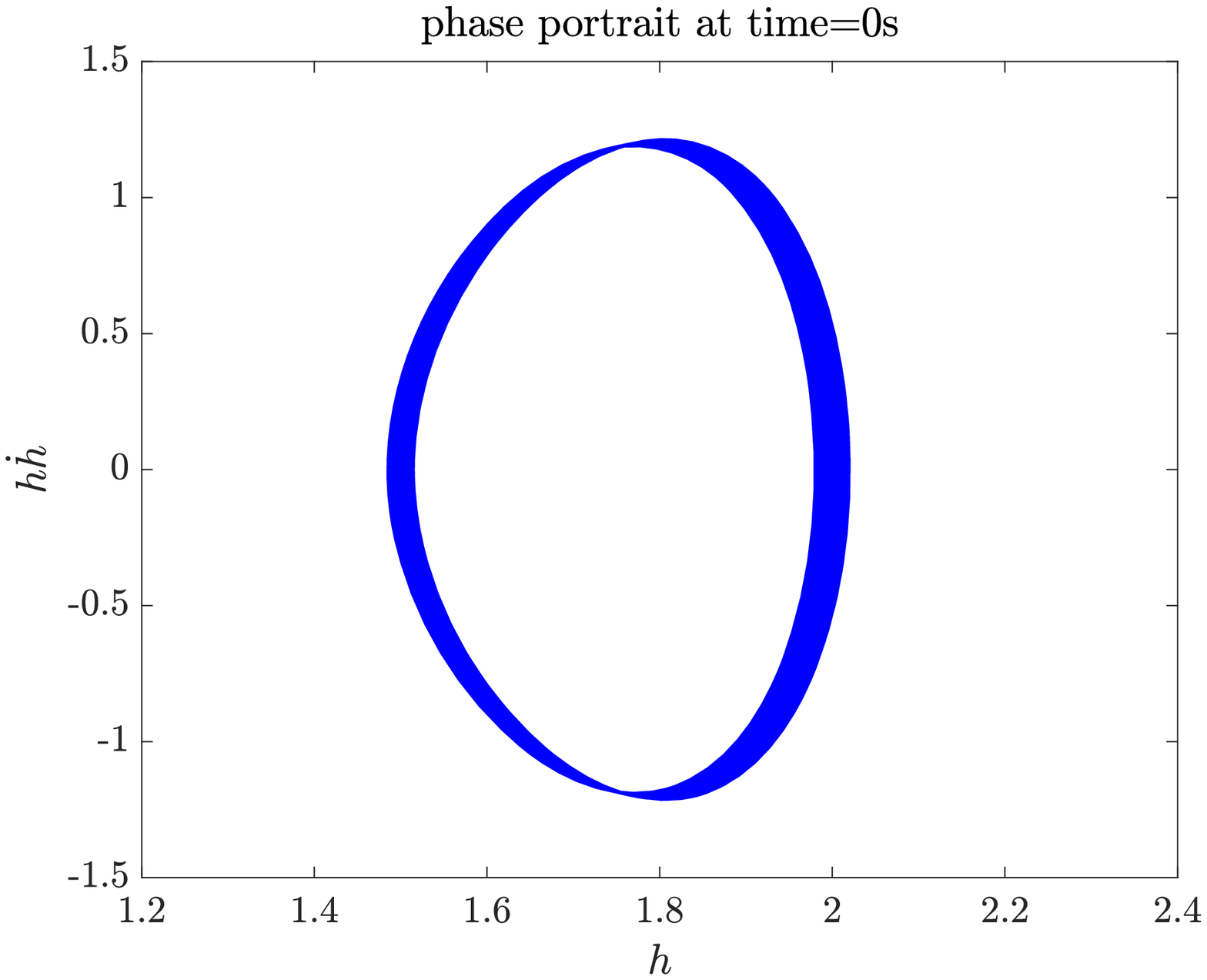}
\end{minipage}
\end{tabular}

\vskip 0.1in
\begin{tabular}{cc}
\begin{minipage}{2.6in}
\hskip 0.1in
\includegraphics[width=2.3in]{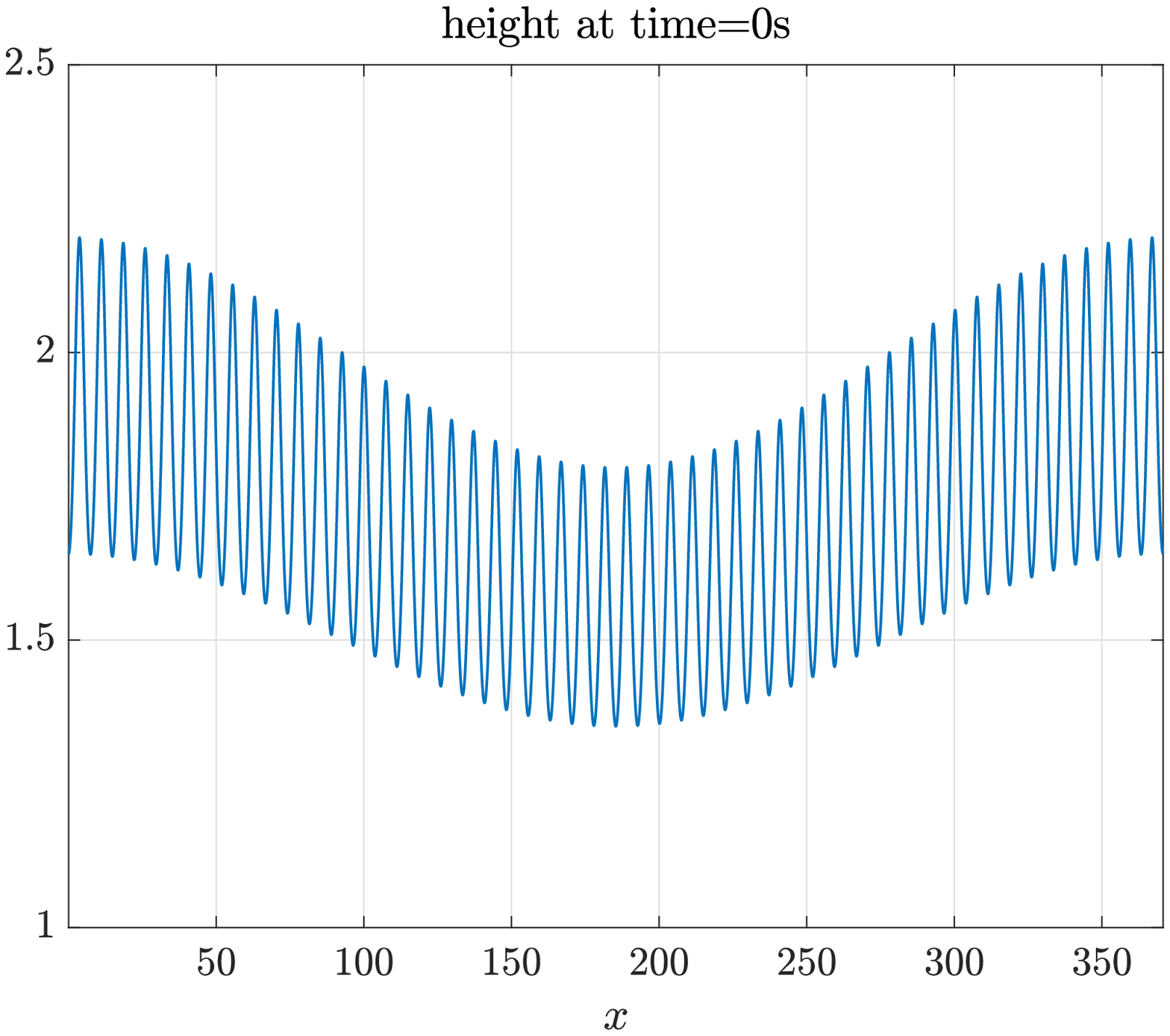}
\end{minipage}
&
\begin{minipage}{2.6in}
\includegraphics[width=2.5in]{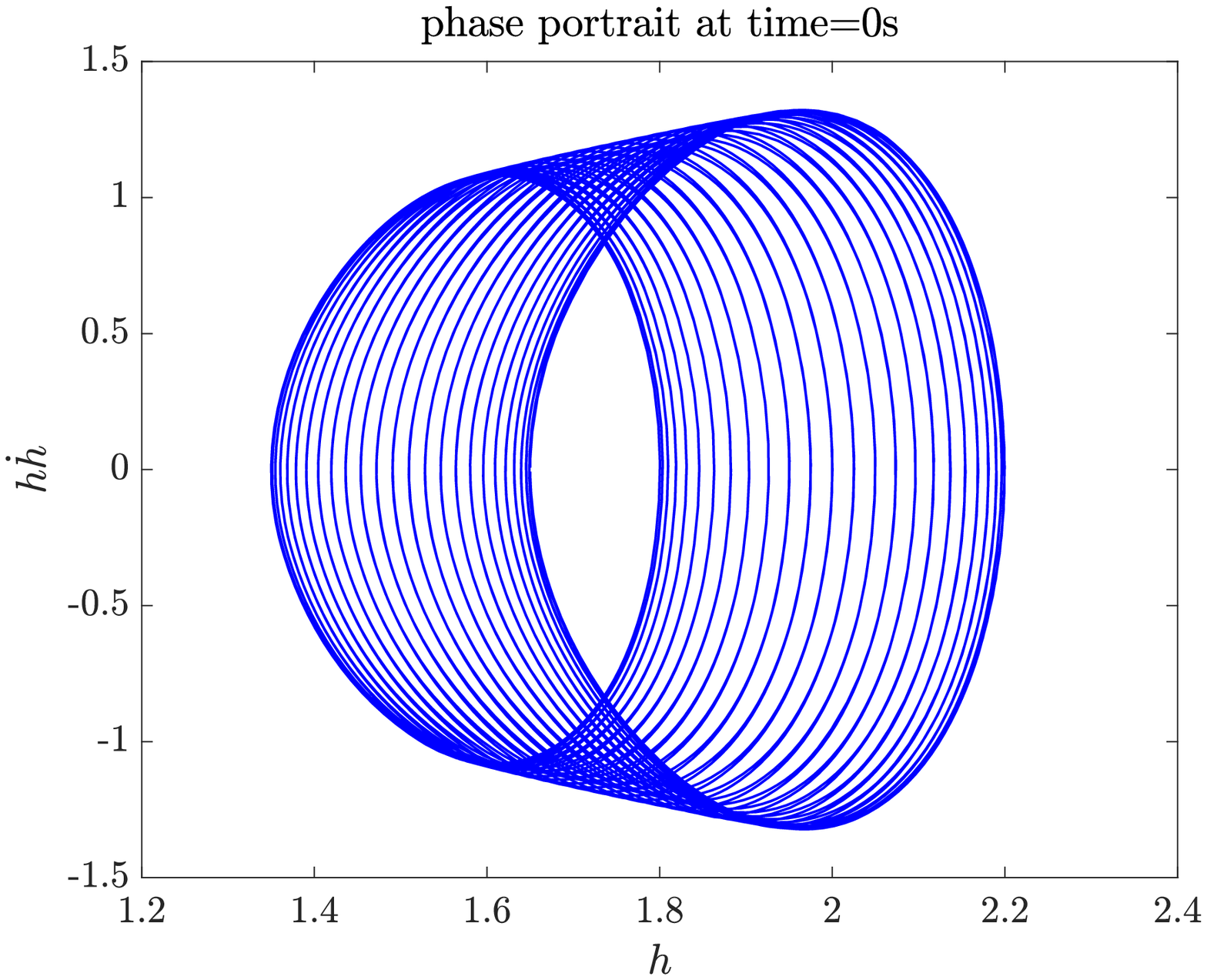}
\end{minipage}

\end{tabular}
\end{center}
\vskip -0.1in
\caption{
    The initial conditions for the modulational test of periodic
    solutions of SGN model;
    water height $h$ is shown on the left,
    and the phase portrait graphed in the $(h, h \dot{h})$
    plane is shown on the right.
    Three different perturbation amplitides, \ie,
    $a=10^{-3}$ (first row),
    $a=10^{-2}$ (second row), and
    $a=10^{-1}$ (third row),
    are considered here with $N=50$.
         }
\label{fig:sgn-mod-t0}
\end{figure}

In the numerical simulations of the SGN model performed below,
we take $N=50$, and perturbation
amplitudes $a=10^{-j}$ for $j=1,2,3$ in the runs.
We show the initial condition of the test in 
Fig.~\ref{fig:sgn-mod-t0}, and the computed solutions
at four different times $t= 200$, $400$, $800$,
$1200$ $s$ in Fig.~\ref{fig:sgn-mod},
where both the water height and the phase portrait in the 
$(h, h\dot{h})$-plane are present. The choice of $h\dot h$ variable is natural, 
because on travelling wave solutions $\displaystyle{h\dot h=m\frac{dh}{d\xi}}$, 
so up to a multiplicative constant, $(h, h\dot{h})$-plane is nothing than the classical 
phase space $\left(h, \frac{dh}{d\xi} \right )$.  
The periodic wave remains  stable when the smaller values
$a = 10^{-2}$ and $10^{-3}$ are used. To see the limit of linear stability, we have also taken  a large amplitude perturbation ($a=10^{-1}$). 
The periodic wave train becomes unstable : we are too far from the classical 
``small perturbation analysis". 
The numerical results are obtained using a hyperbolic-elliptic
splitting method proposed by the authors~\cite{Gavrilyuk_2018}
with $400$ meshes for each wave length.

\begin{figure}[!h]
\begin{center}
\begin{tabular}{cc}
\begin{minipage}{2.6in}
\hskip 0.1in
\includegraphics[width=2.3in]{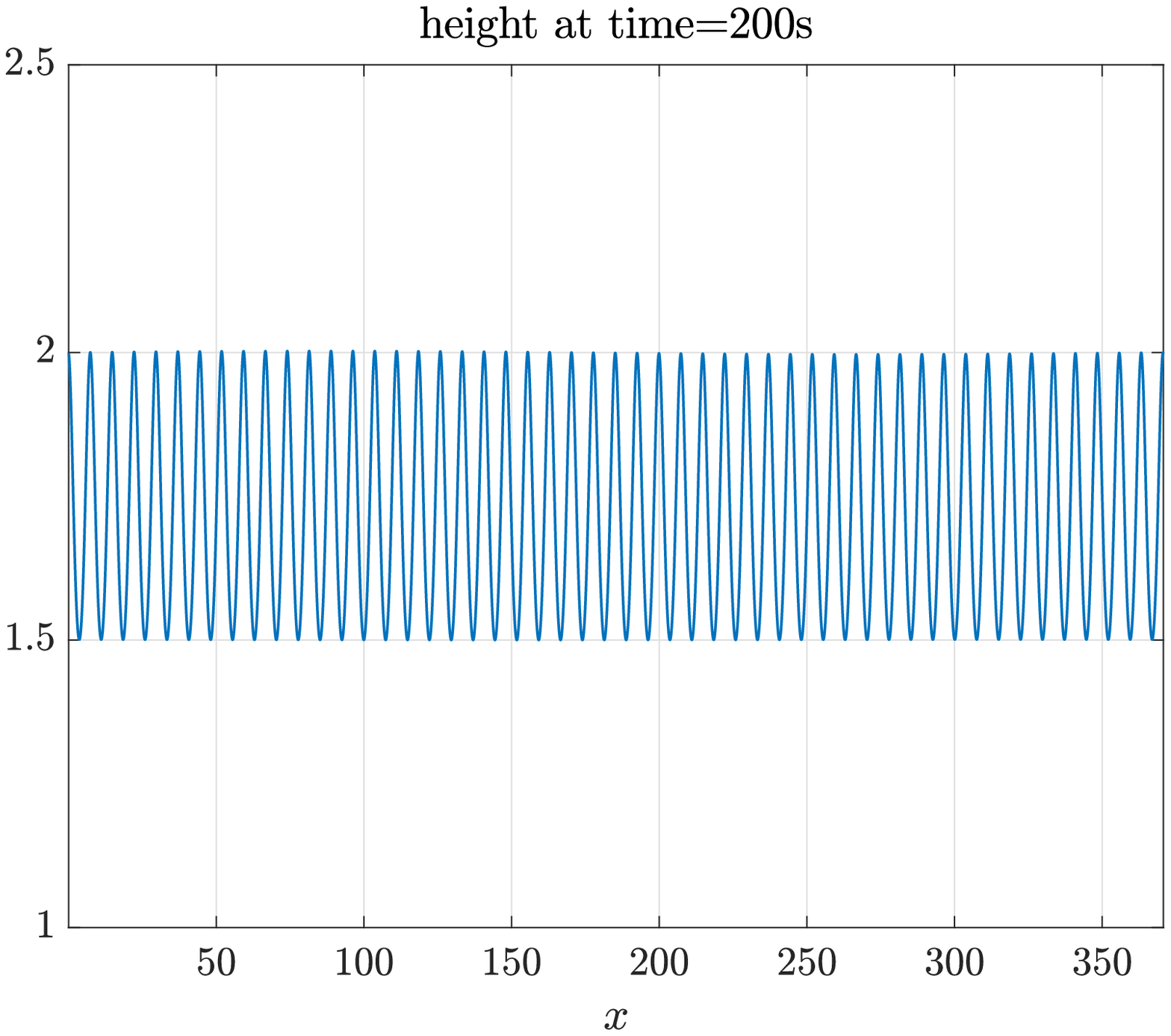}
\end{minipage}
&
\begin{minipage}{2.6in}
\includegraphics[width=2.5in]{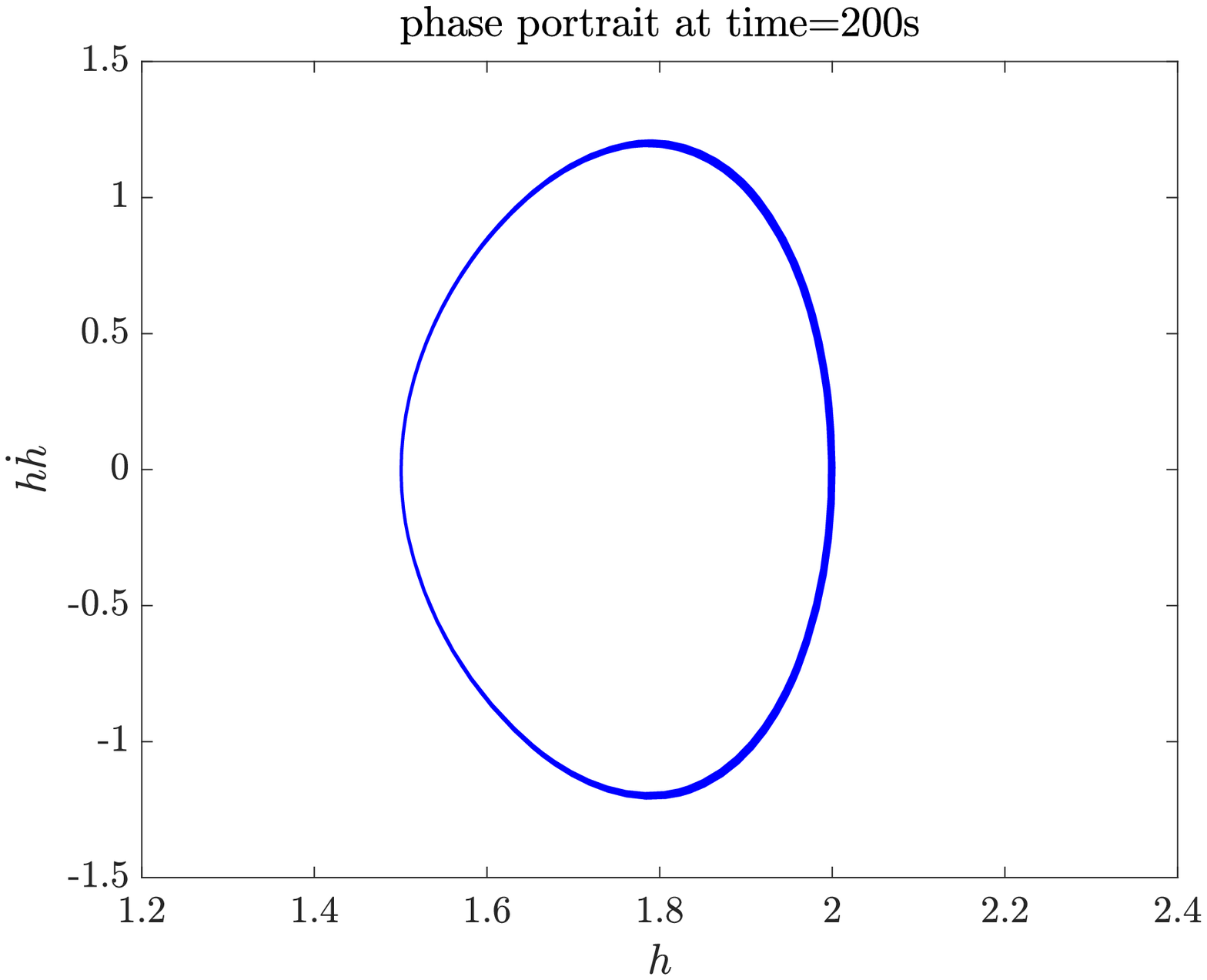}
\end{minipage}
\end{tabular}
\vskip 0.1in
\begin{tabular}{cc}
\begin{minipage}{2.6in}
\hskip 0.1in
\includegraphics[width=2.3in]{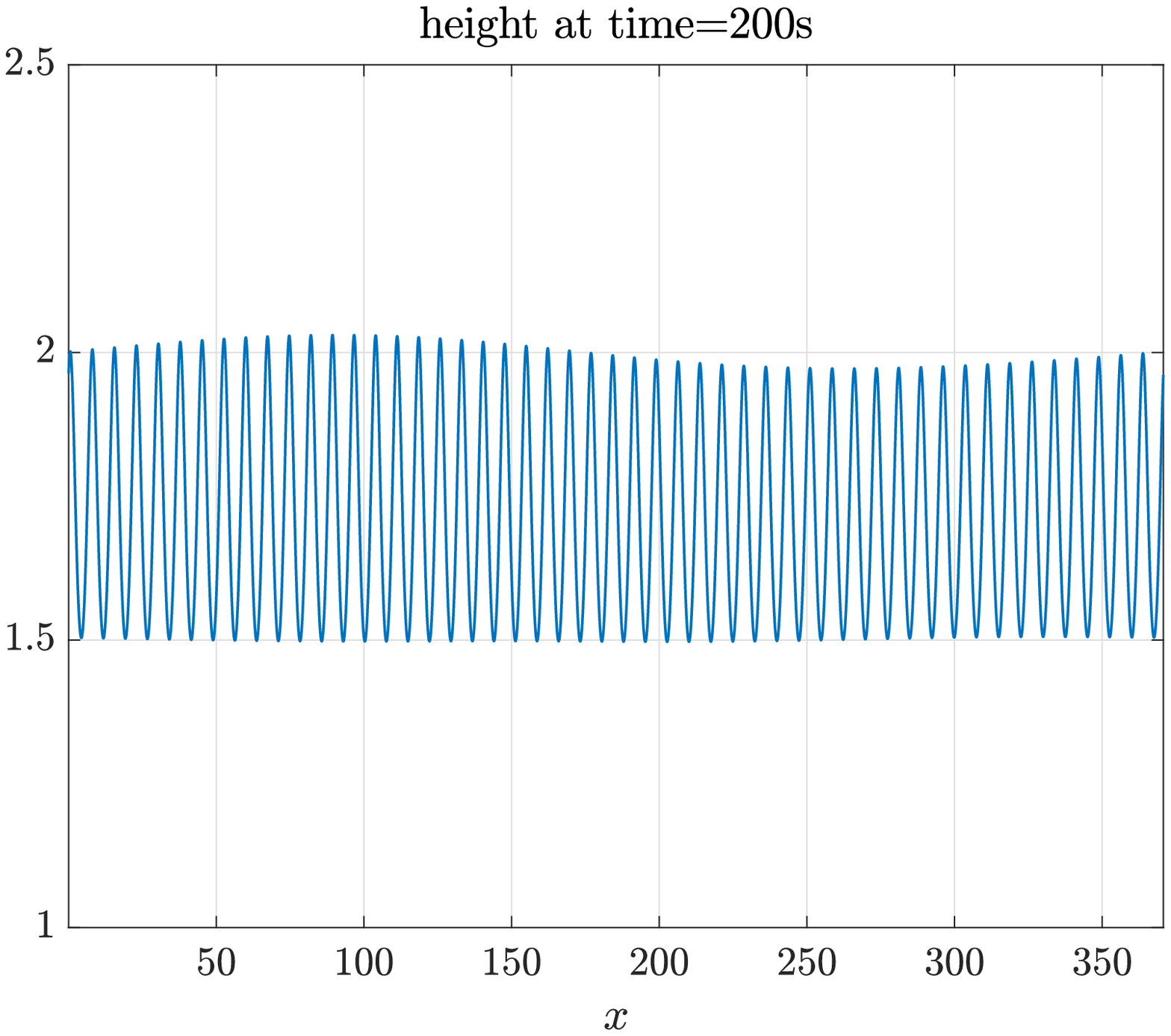}
\end{minipage}
&
\begin{minipage}{2.6in}
\includegraphics[width=2.5in]{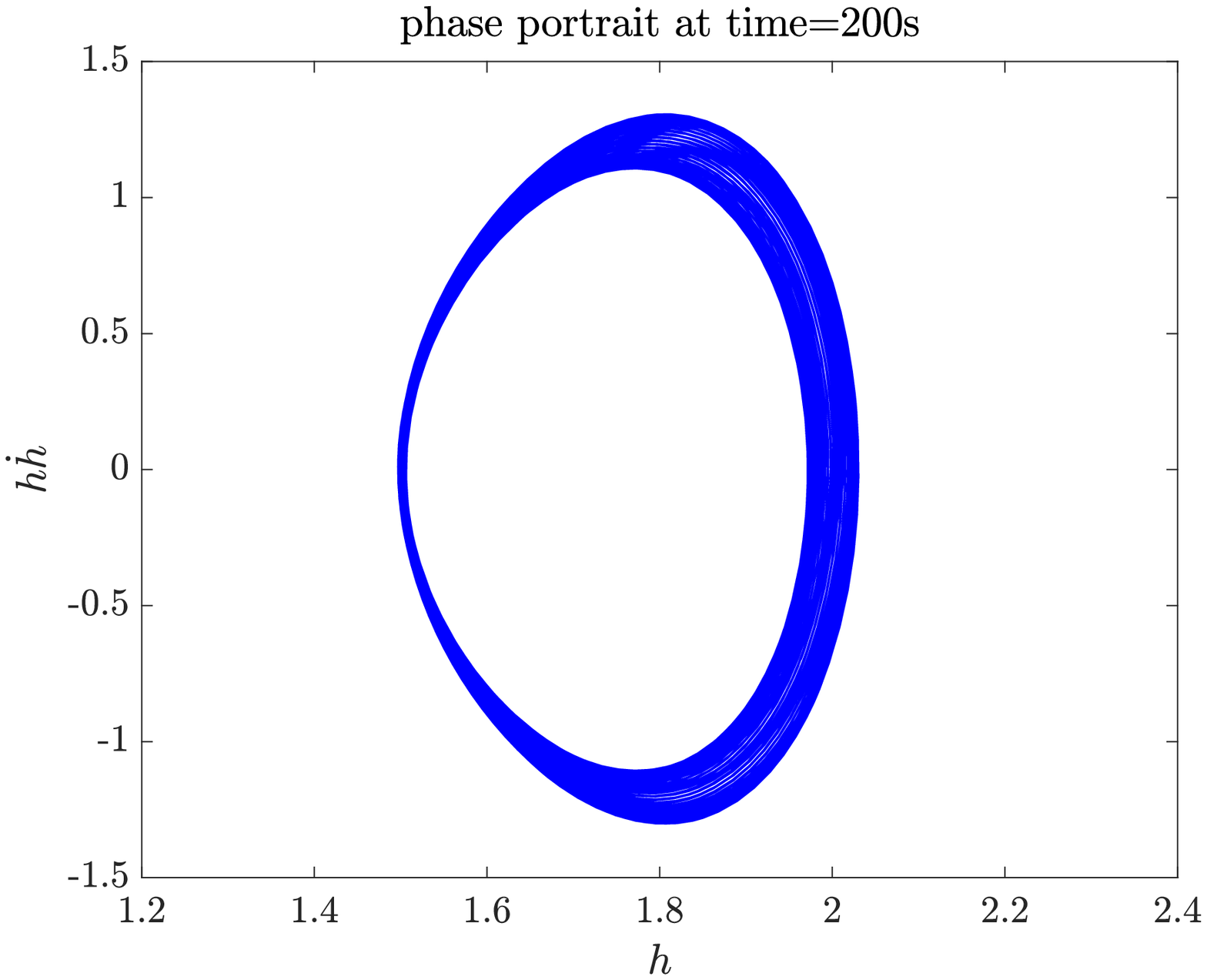}
\end{minipage}
\end{tabular}

\vskip 0.1in
\begin{tabular}{cc}
\begin{minipage}{2.6in}
\hskip 0.1in
\includegraphics[width=2.3in]{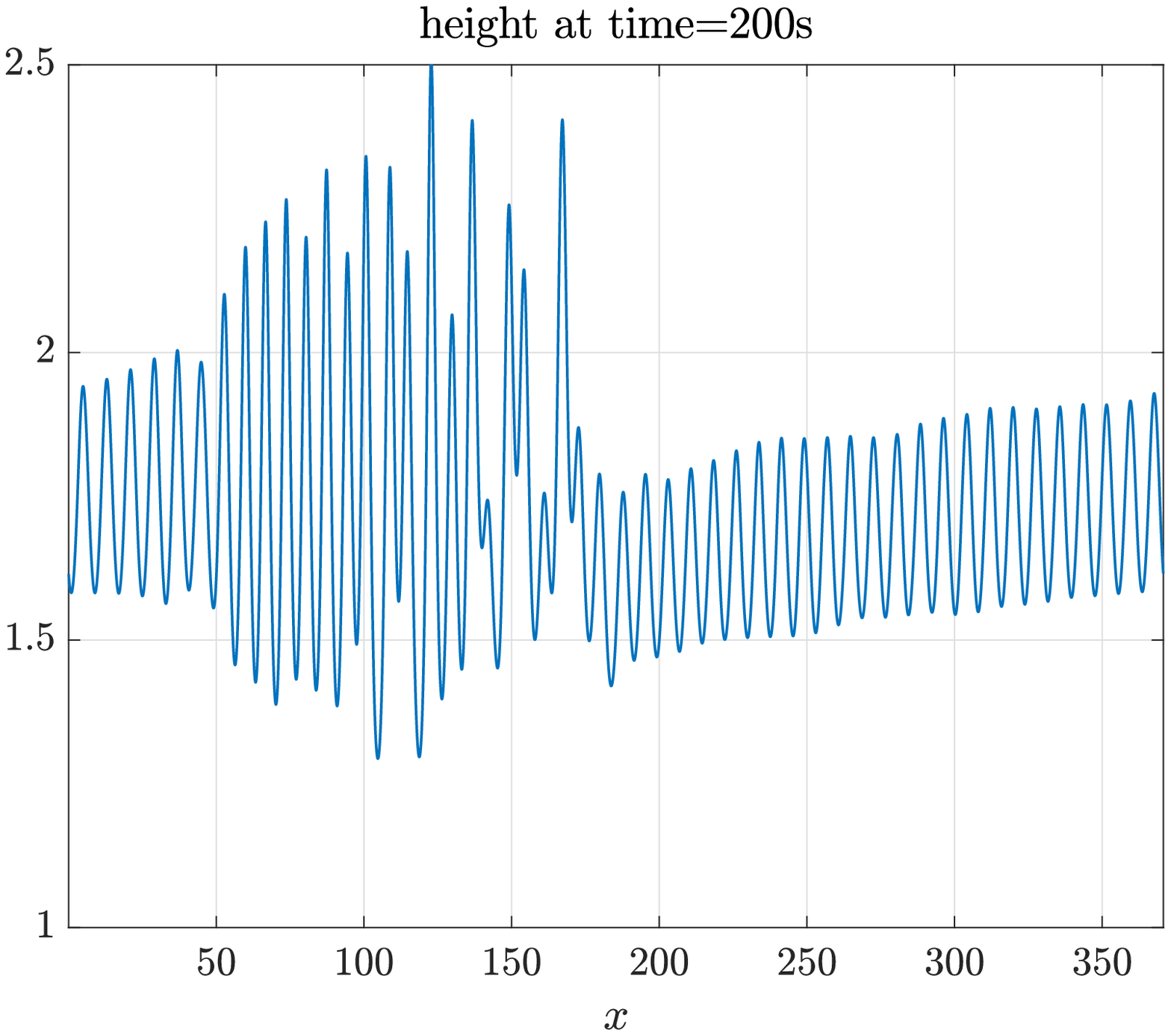}
\end{minipage}
&
\begin{minipage}{2.6in}
\includegraphics[width=2.5in]{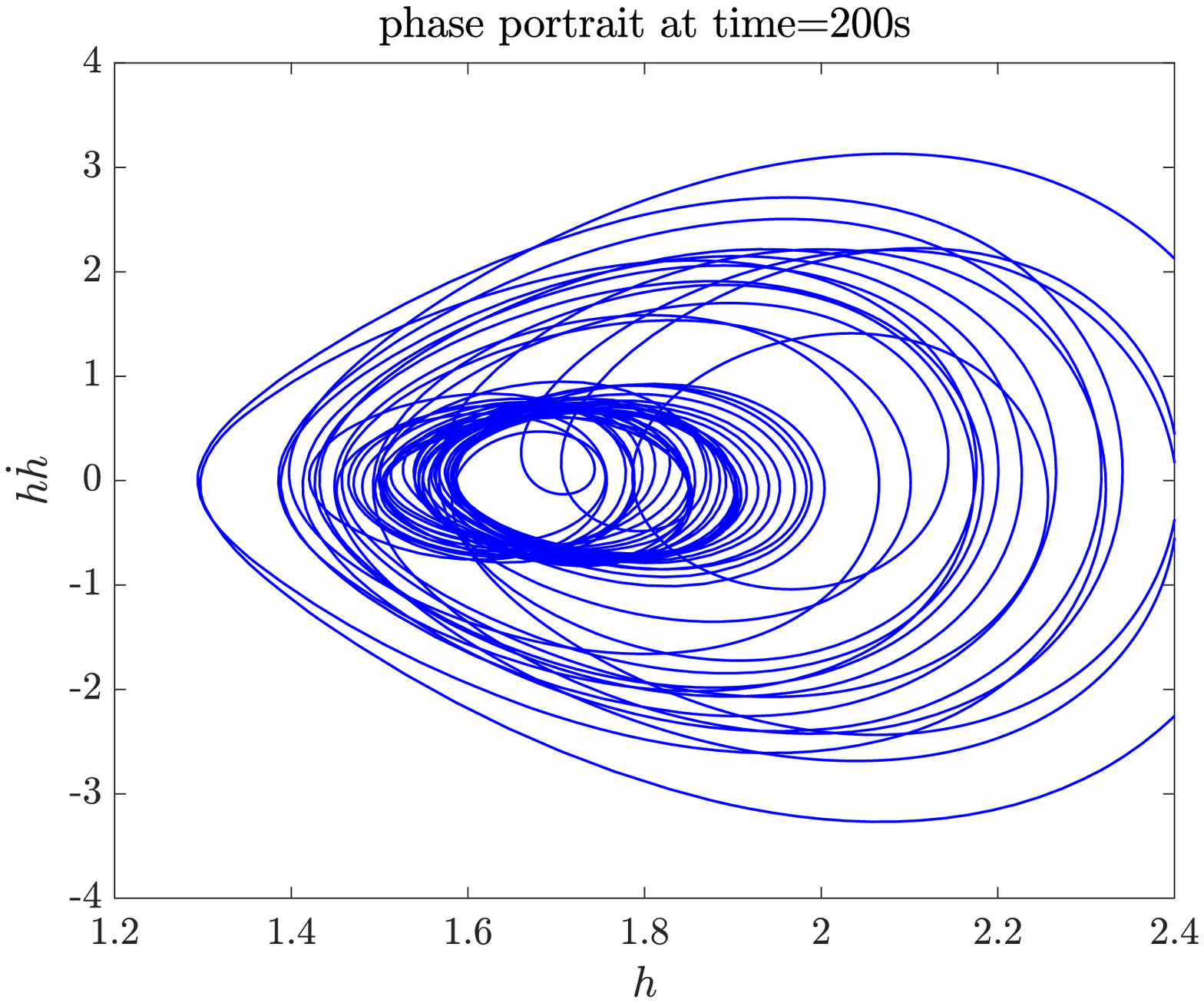}
\end{minipage}

\end{tabular}
\end{center}
\vskip -0.1in
\caption{
    Numerical results for a modulational stability test of periodic
    waves shown for four  time instants : $t= 200$, $400$, $800$,
    $1200$ $s$. The graphs are displayed in the same manner as 
    in Fig.~\ref{fig:sgn-mod-t0}.
         }
\label{fig:sgn-mod}
\end{figure}

\begin{figure}[!h]
\setcounter{figure}{4}
\begin{center}
\begin{tabular}{cc}
\begin{minipage}{2.6in}
\hskip 0.1in
\includegraphics[width=2.3in]{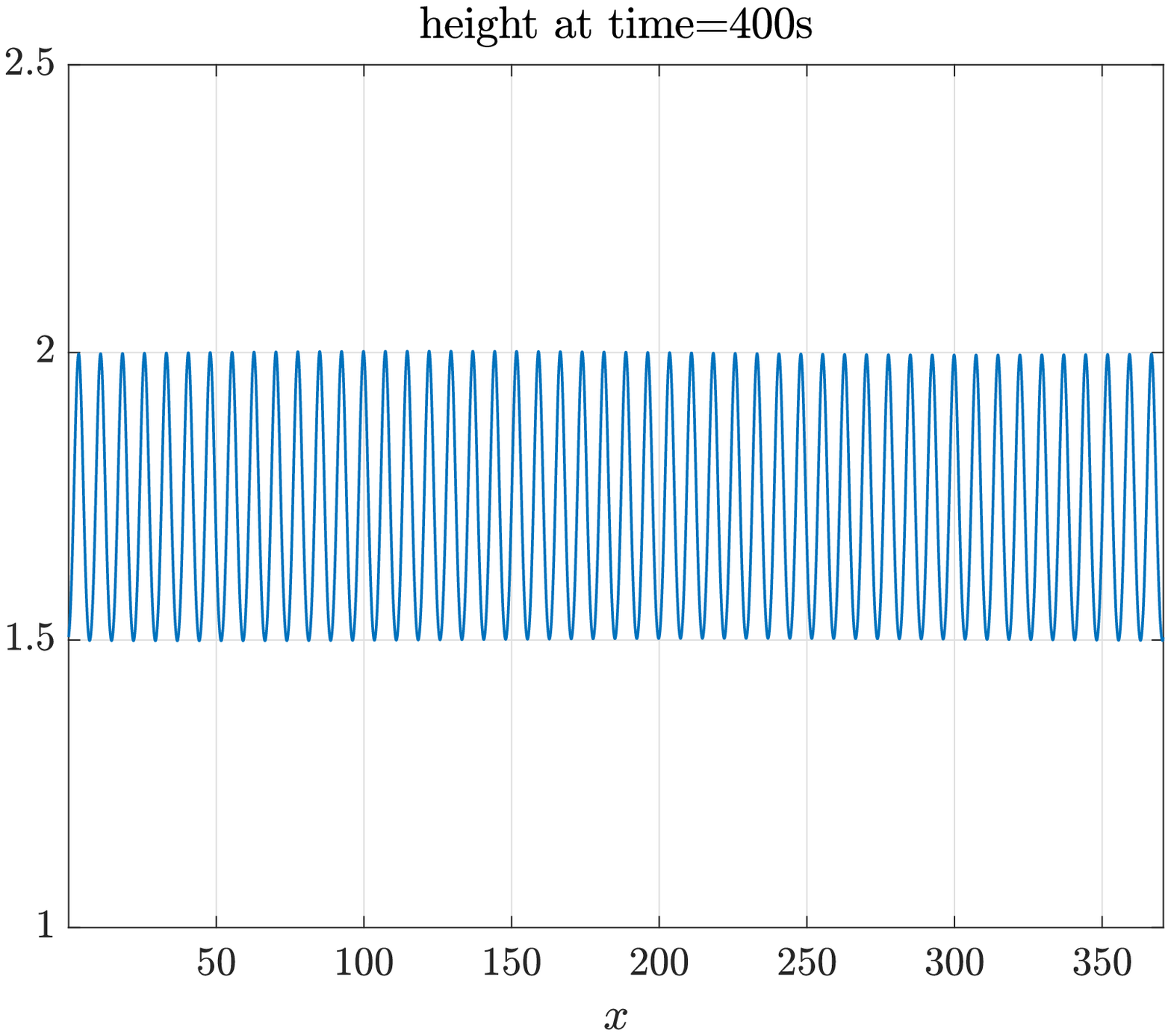}
\end{minipage}
&
\begin{minipage}{2.6in}
\includegraphics[width=2.5in]{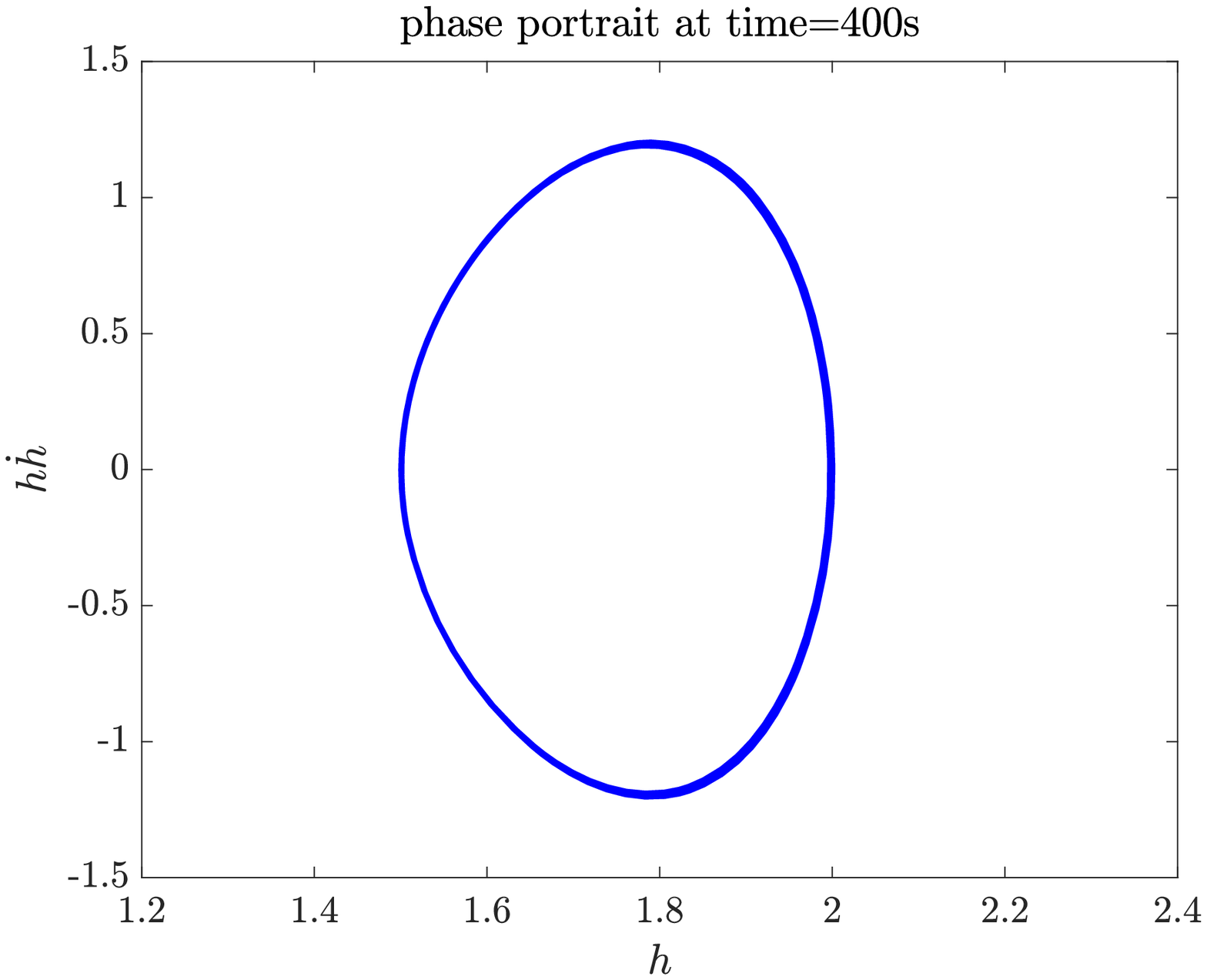}
\end{minipage}
\end{tabular}
\vskip 0.1in
\begin{tabular}{cc}
\begin{minipage}{2.6in}
\hskip 0.1in
\includegraphics[width=2.3in]{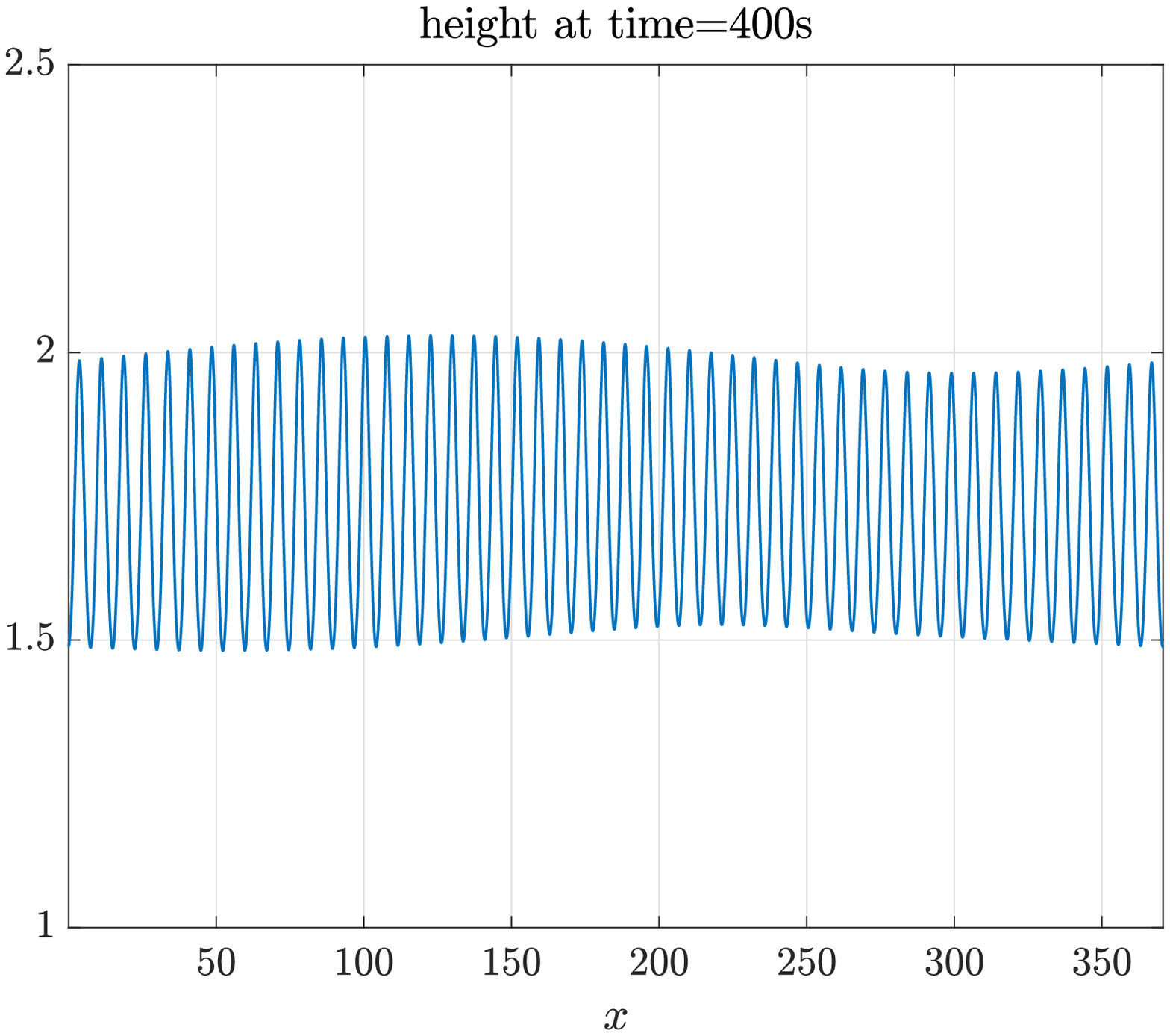}
\end{minipage}
&
\begin{minipage}{2.6in}
\includegraphics[width=2.5in]{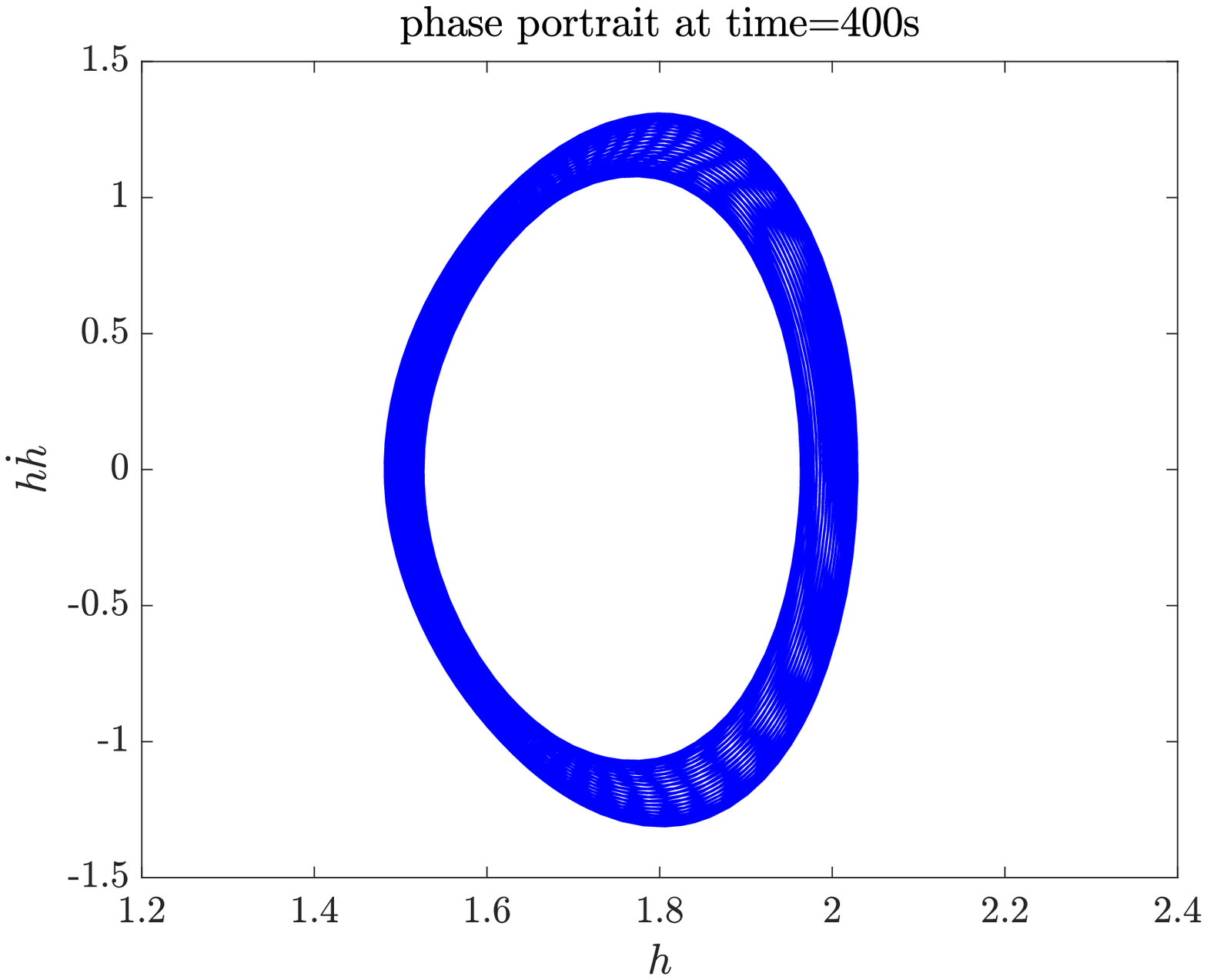}
\end{minipage}
\end{tabular}

\vskip 0.1in
\begin{tabular}{cc}
\begin{minipage}{2.6in}
\hskip 0.1in
\includegraphics[width=2.3in]{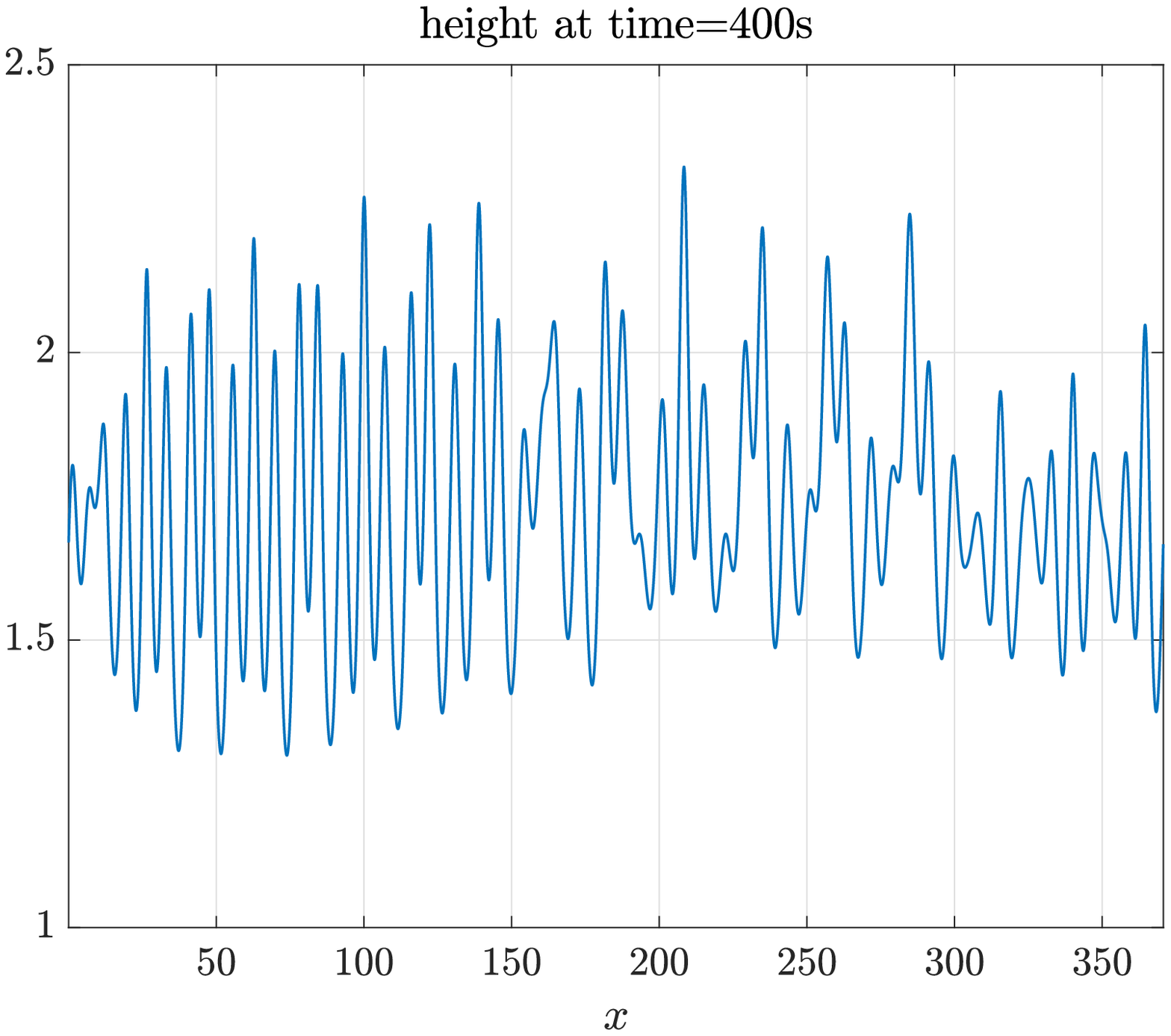}
\end{minipage}
&
\begin{minipage}{2.6in}
\includegraphics[width=2.5in]{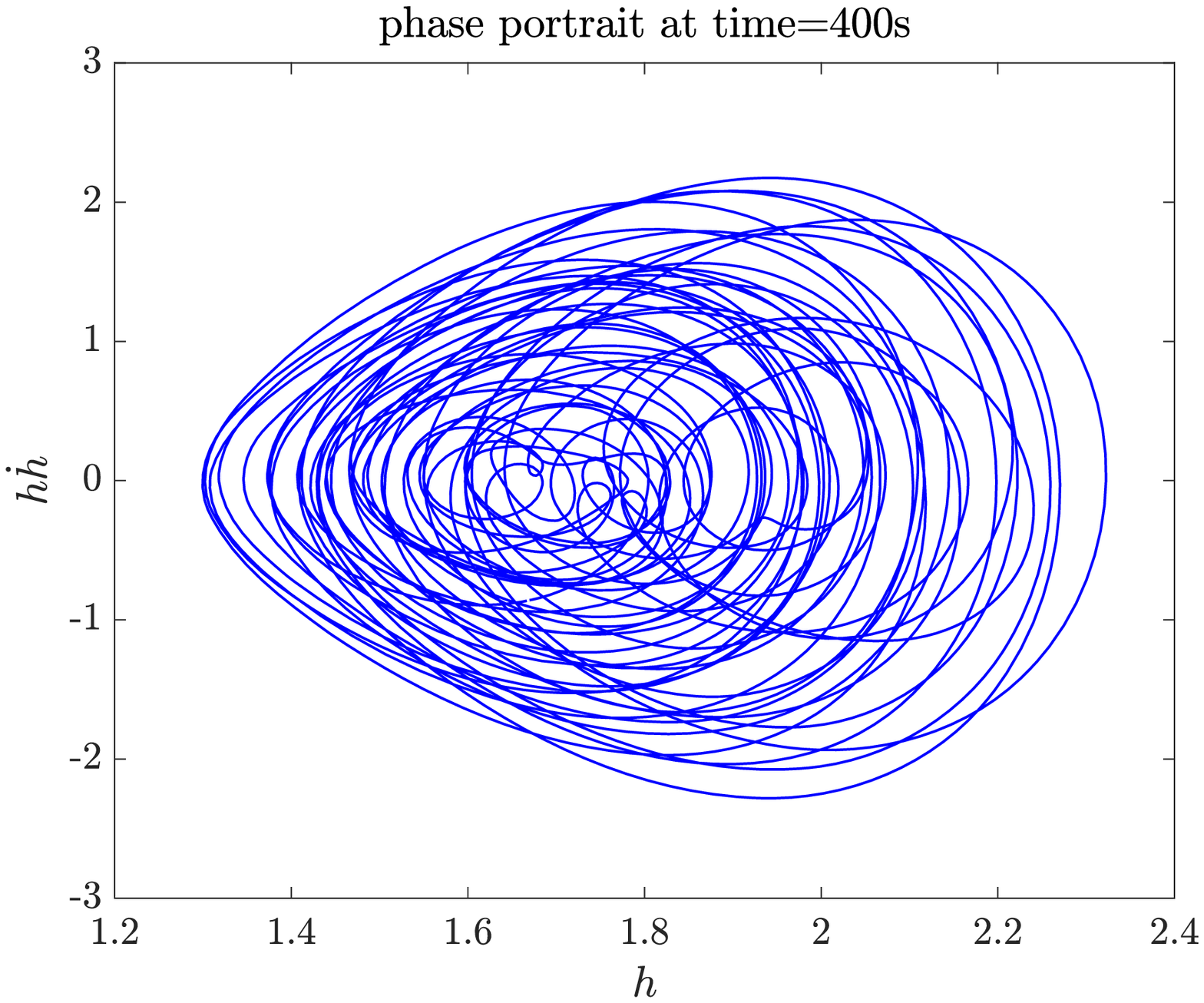}
\end{minipage}

\end{tabular}
\end{center}
\addtocounter{figure}{-1}
\vskip -0.1in
\caption{
       Continued.
         }
\end{figure}

\begin{figure}[!h]
\setcounter{figure}{4}
\begin{center}
\begin{tabular}{cc}
\begin{minipage}{2.6in}
\hskip 0.1in
\includegraphics[width=2.3in]{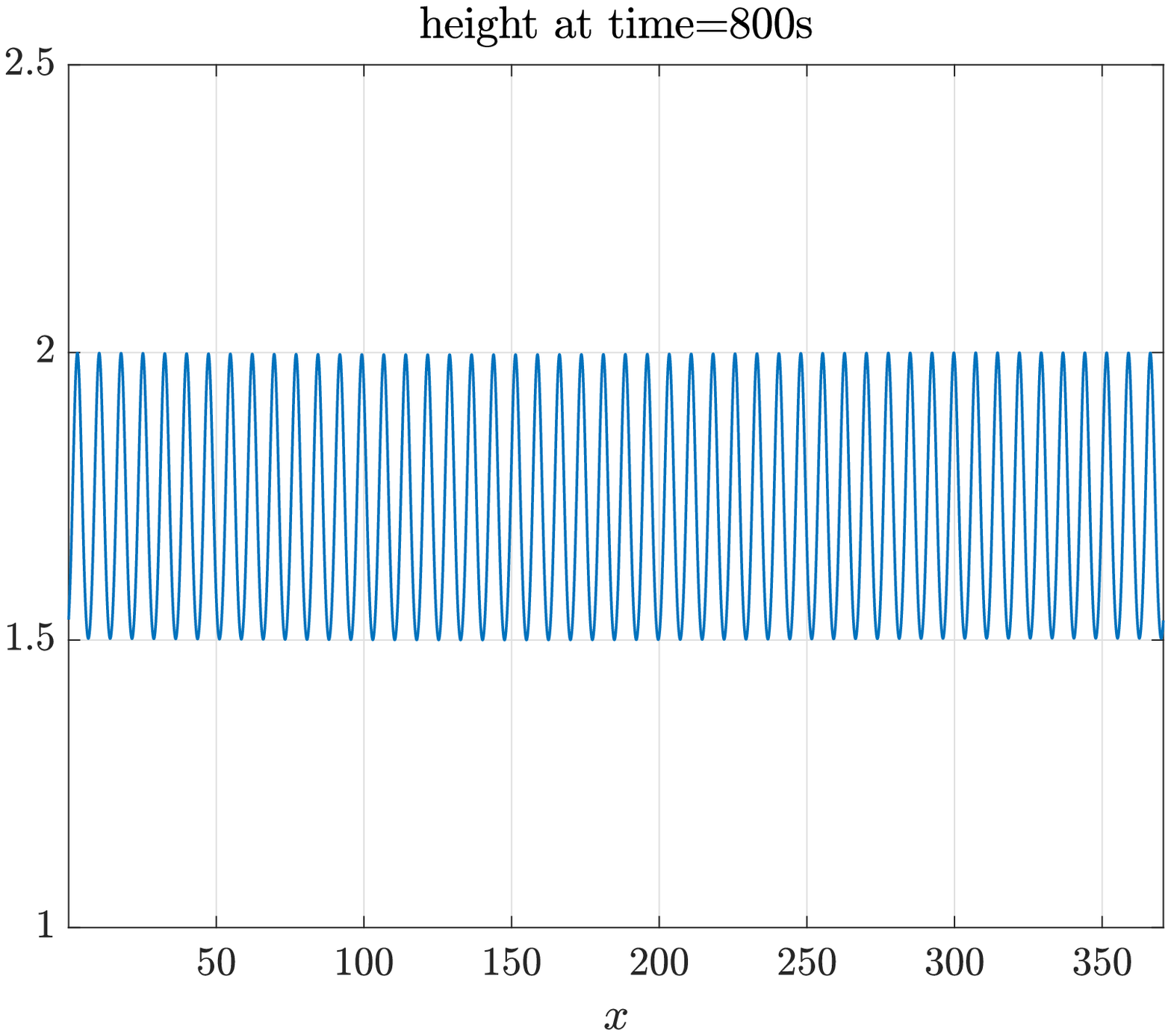}
\end{minipage}
&
\begin{minipage}{2.6in}
\includegraphics[width=2.5in]{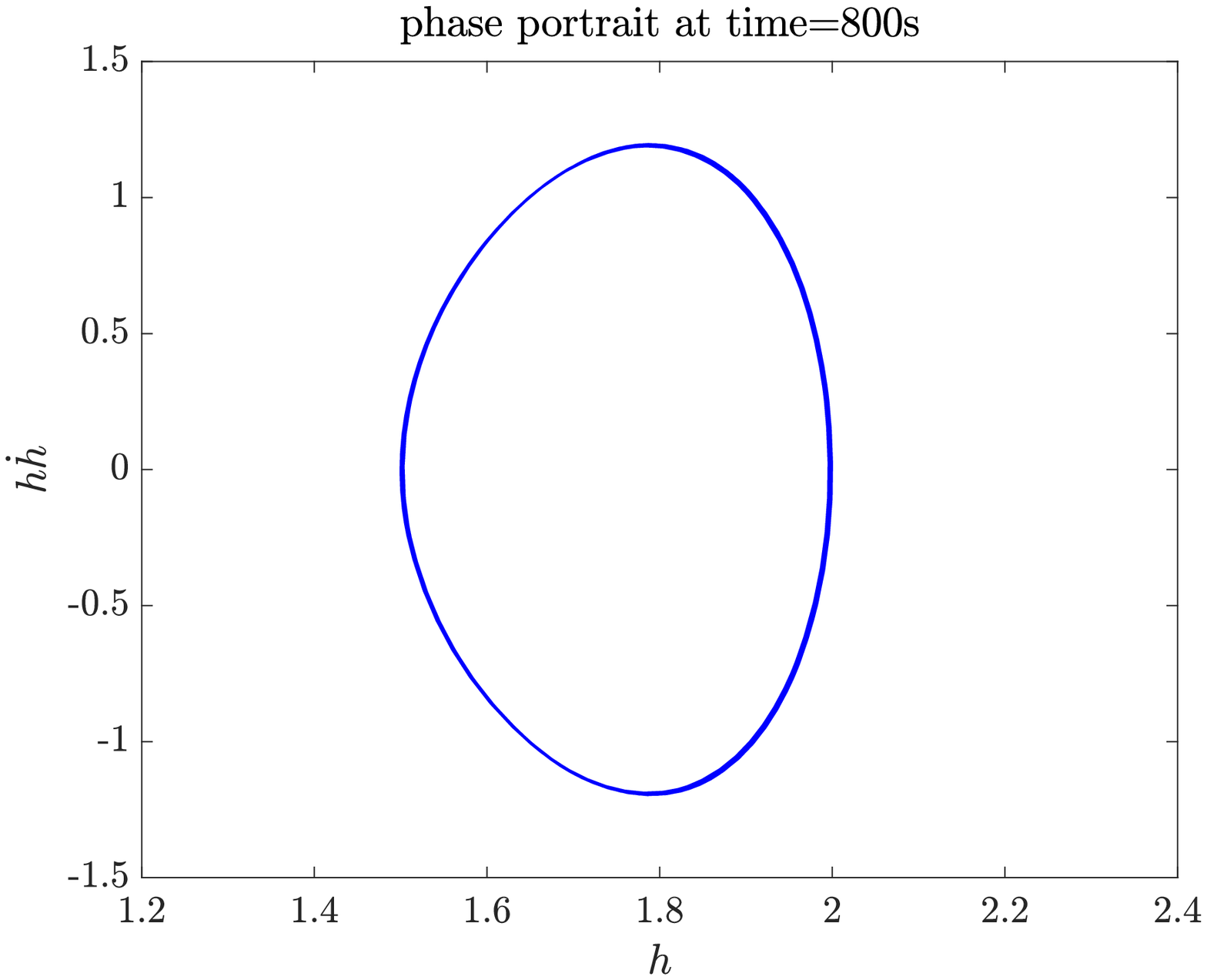}
\end{minipage}
\end{tabular}
\vskip 0.1in
\begin{tabular}{cc}
\begin{minipage}{2.6in}
\hskip 0.1in
\includegraphics[width=2.3in]{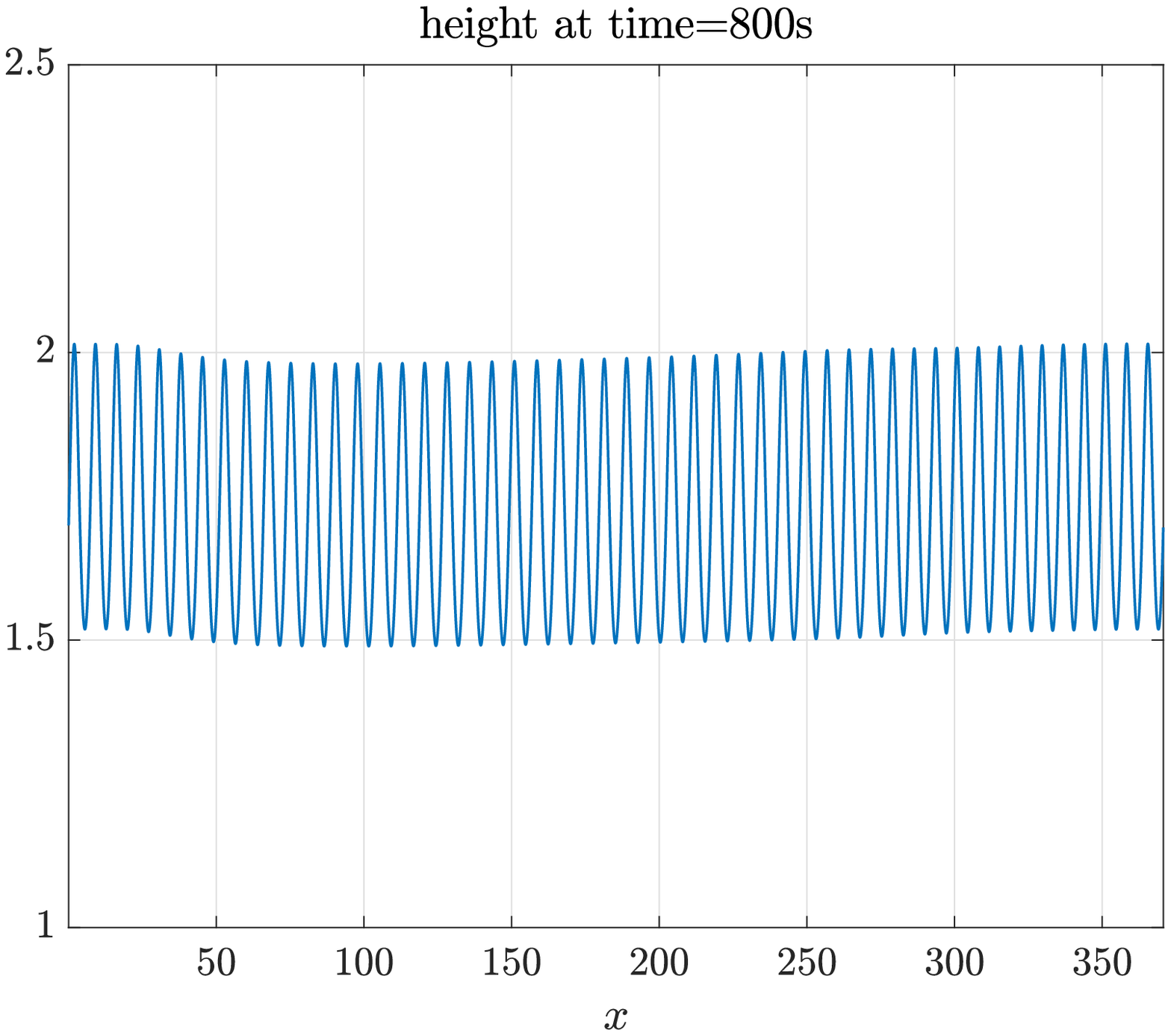}
\end{minipage}
&
\begin{minipage}{2.6in}
\includegraphics[width=2.5in]{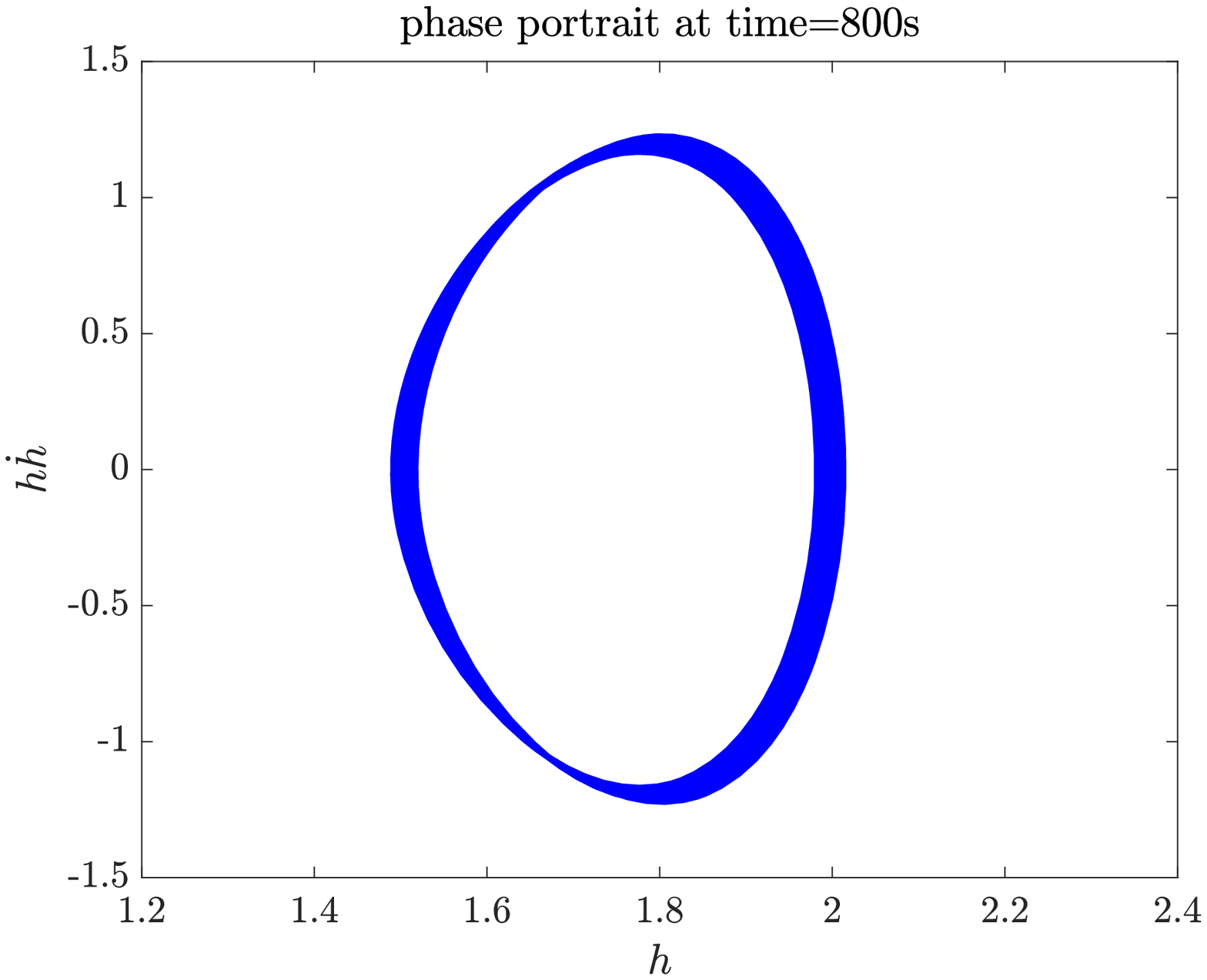}
\end{minipage}
\end{tabular}

\vskip 0.1in
\begin{tabular}{cc}
\begin{minipage}{2.6in}
\hskip 0.1in
\includegraphics[width=2.3in]{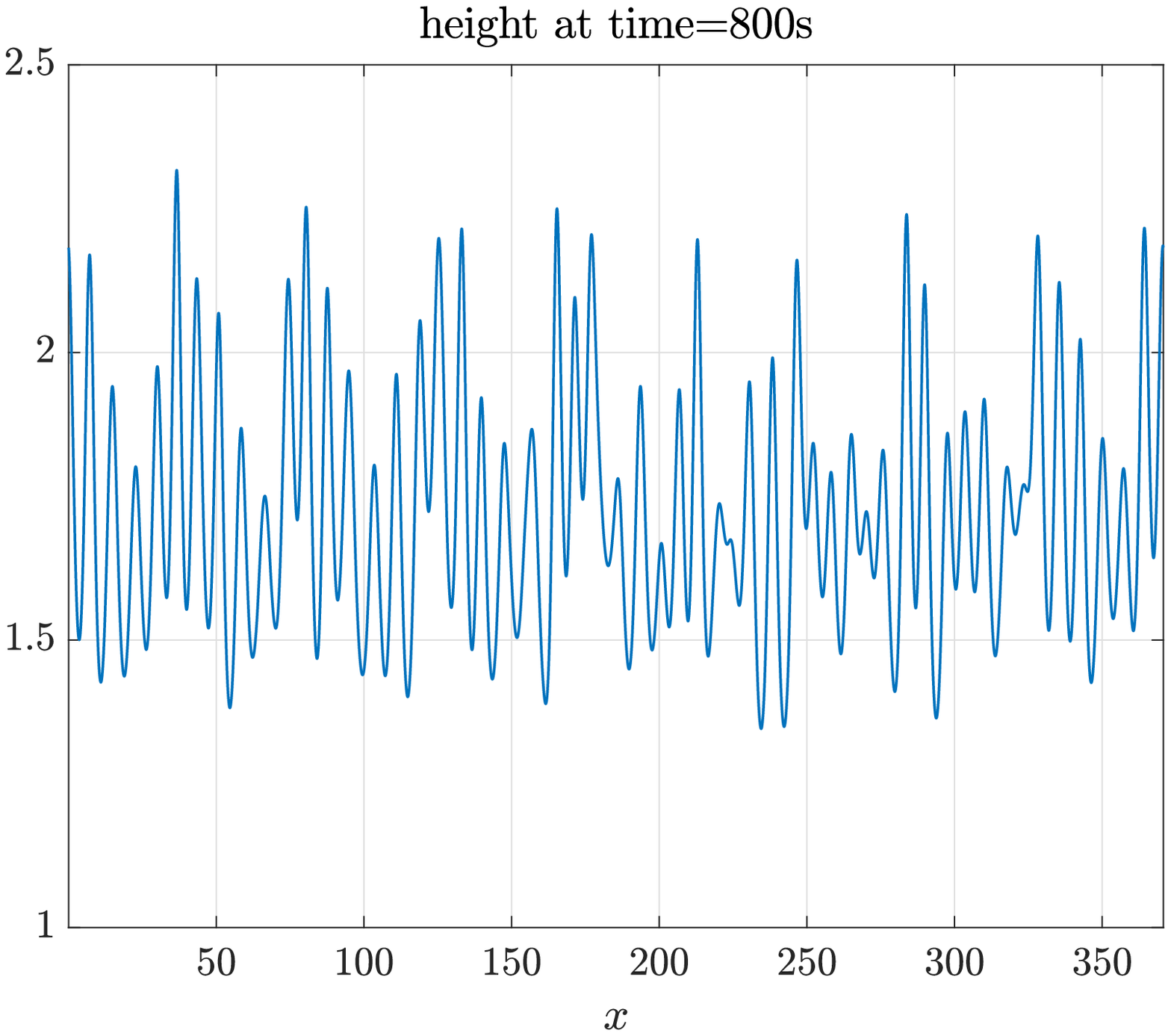}
\end{minipage}
&
\begin{minipage}{2.6in}
\includegraphics[width=2.5in]{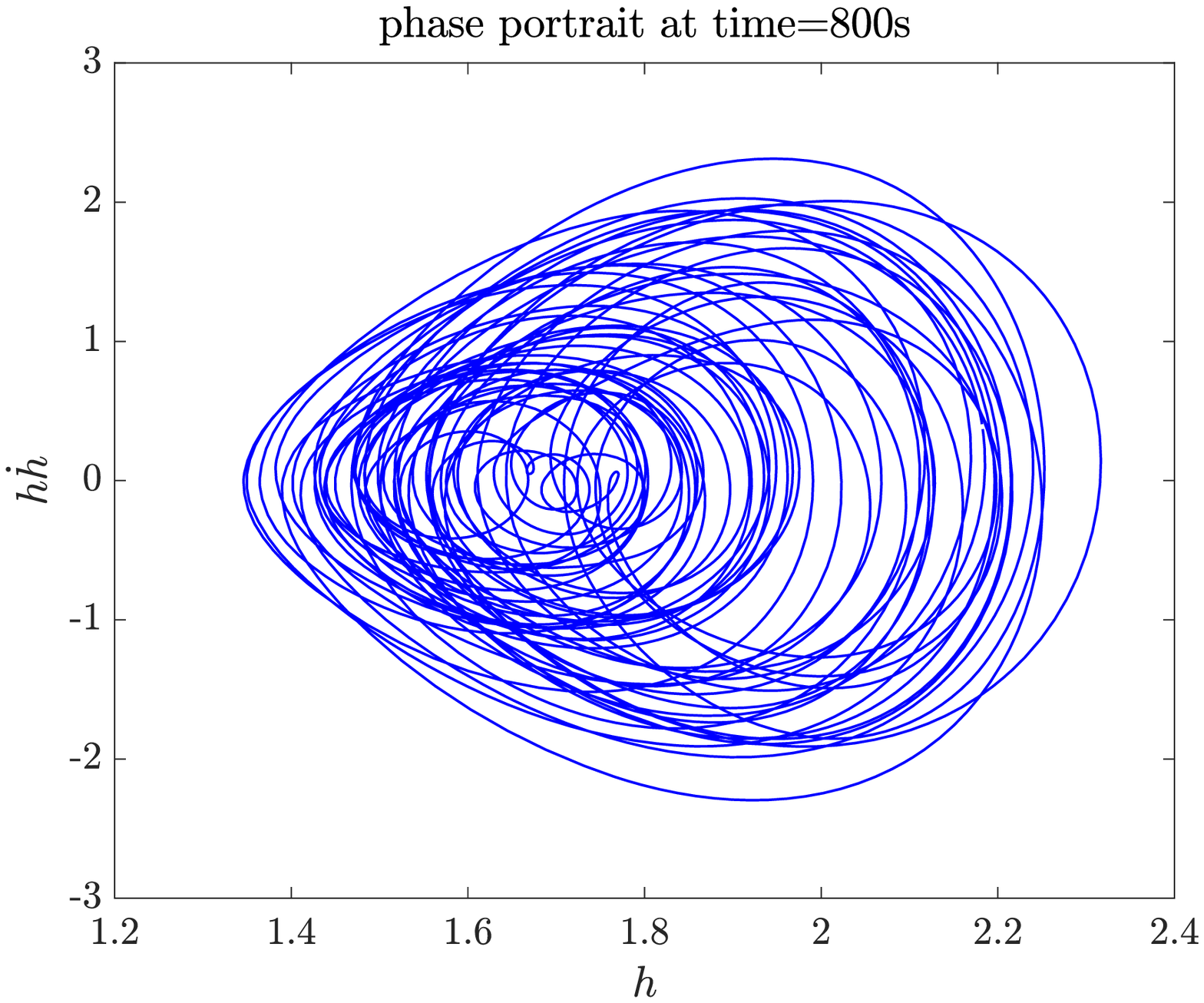}
\end{minipage}

\end{tabular}
\end{center}
\addtocounter{figure}{-1}
\vskip -0.1in
\caption{
     Continued.
         }
\end{figure}

\begin{figure}[!h]
\setcounter{figure}{4}
\begin{center}
\begin{tabular}{cc}
\begin{minipage}{2.6in}
\hskip 0.1in
\includegraphics[width=2.3in]{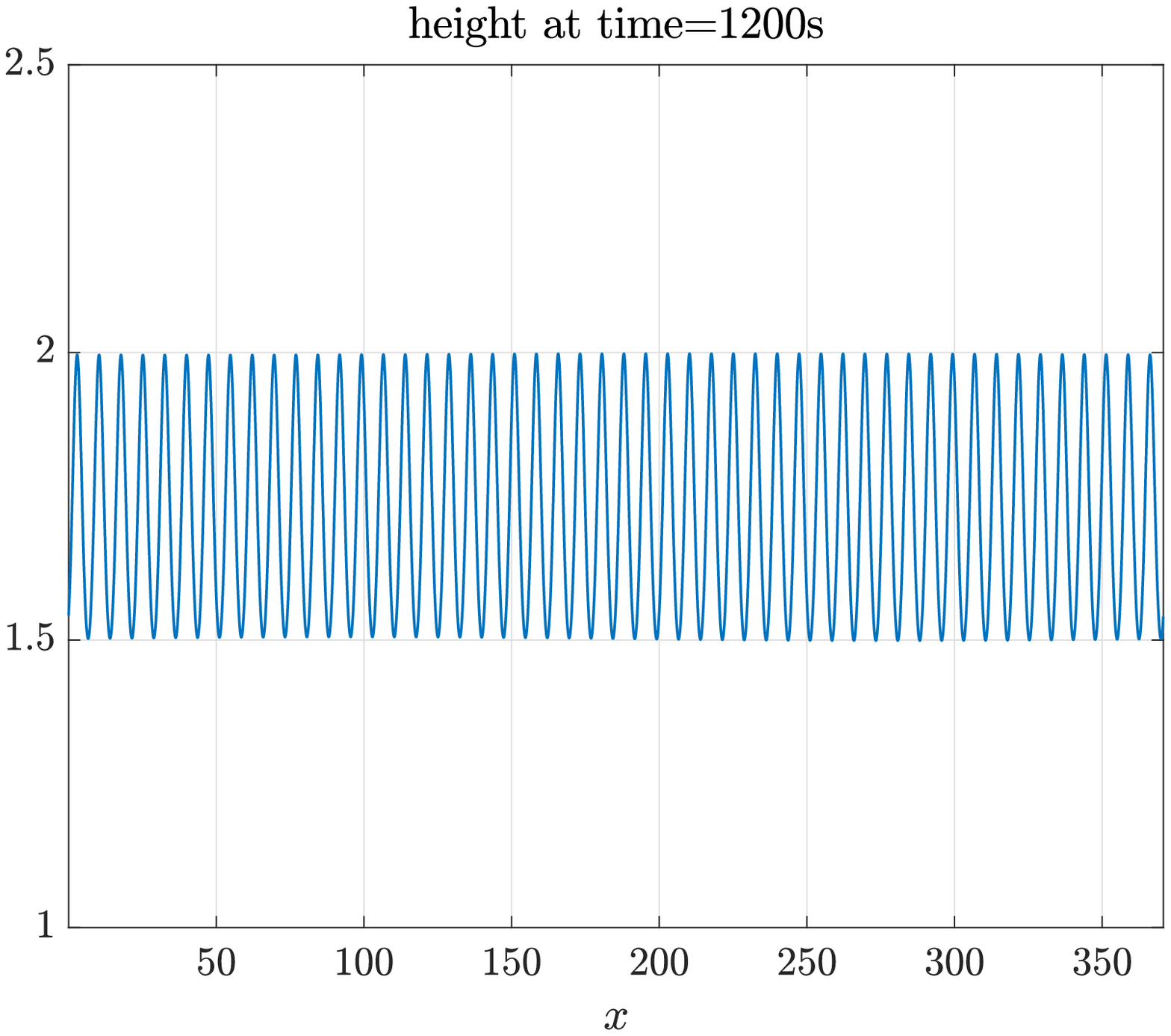}
\end{minipage}
&
\begin{minipage}{2.6in}
\includegraphics[width=2.5in]{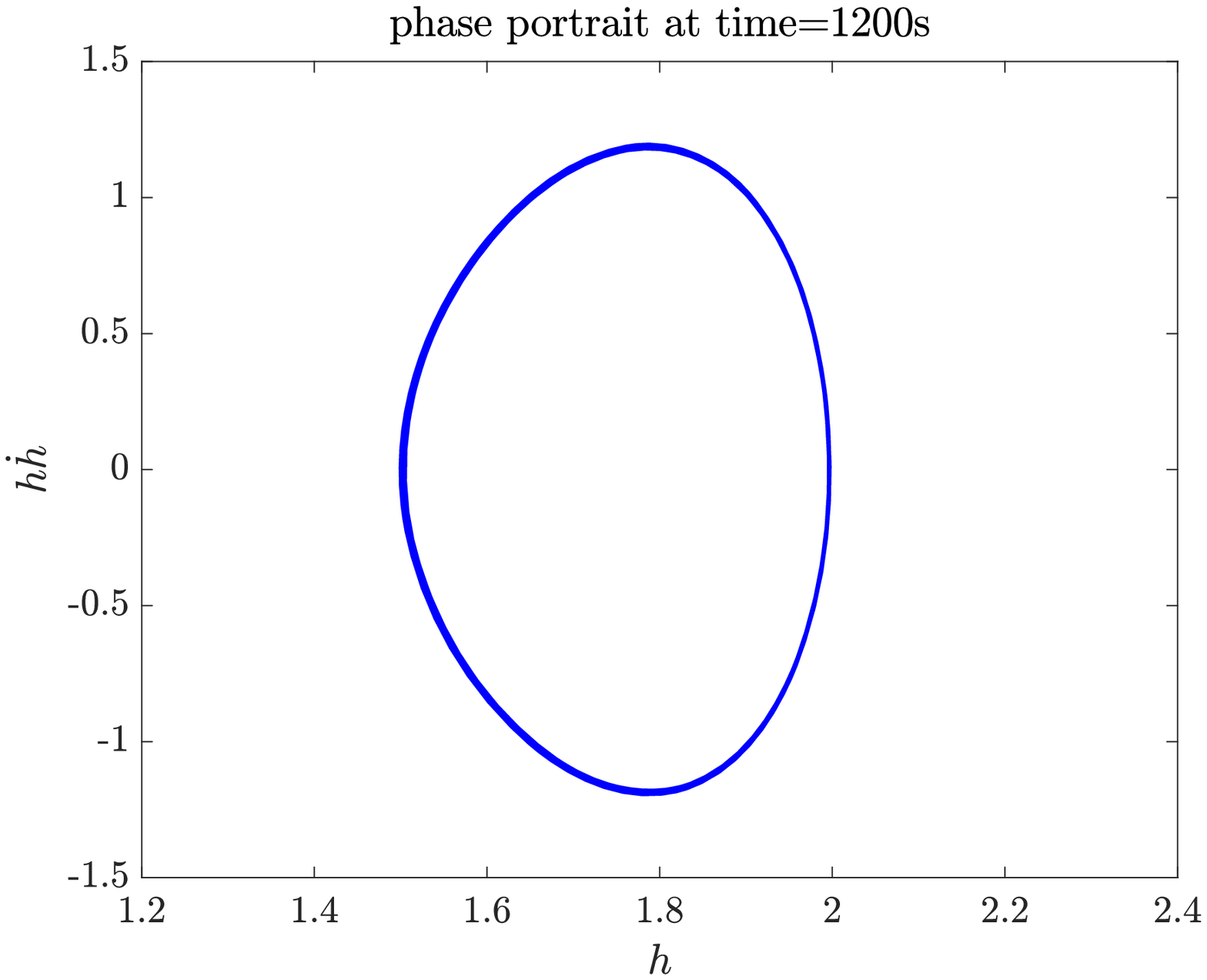}
\end{minipage}
\end{tabular}
\vskip 0.1in
\begin{tabular}{cc}
\begin{minipage}{2.6in}
\hskip 0.1in
\includegraphics[width=2.3in]{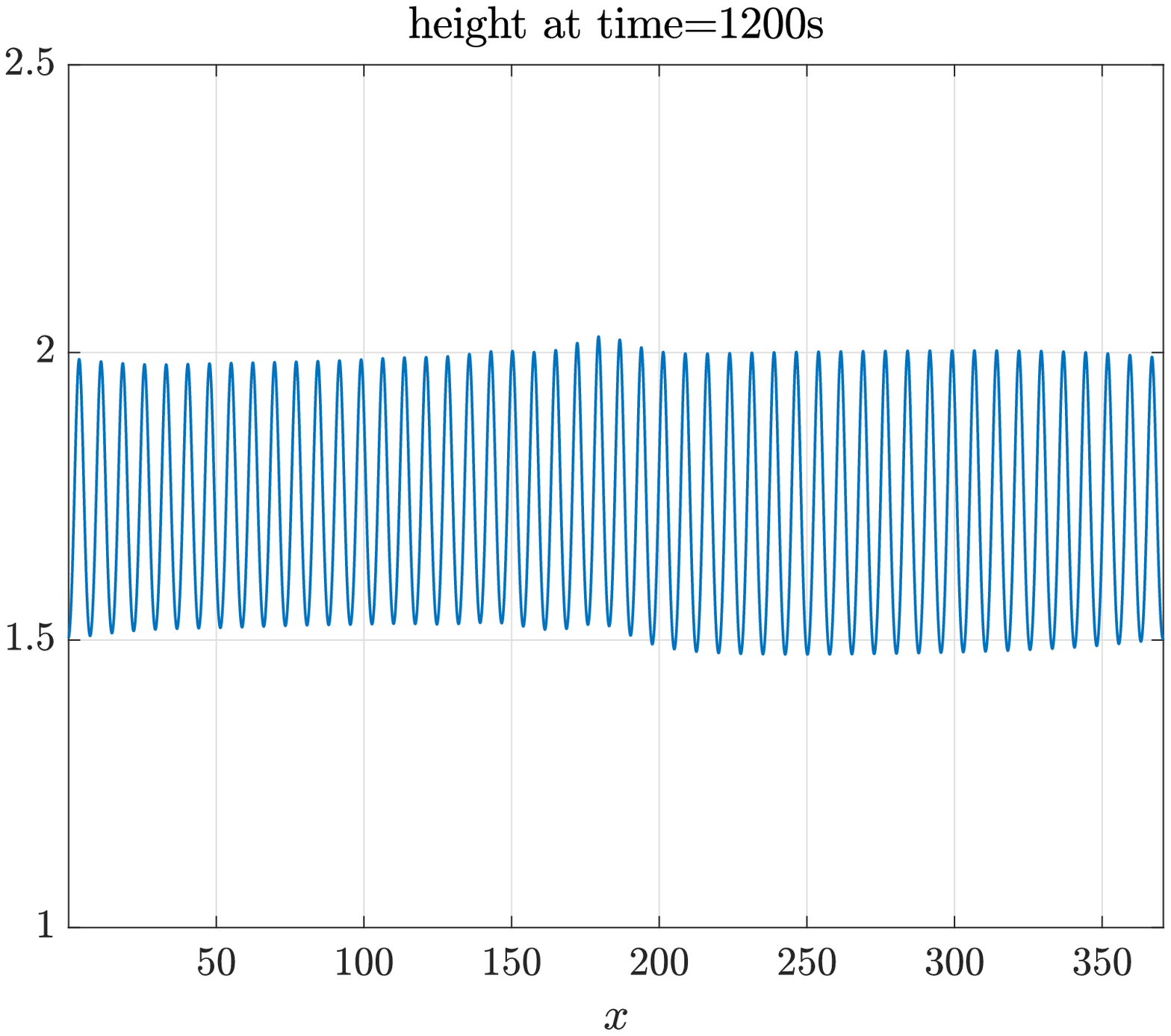}
\end{minipage}
&
\begin{minipage}{2.6in}
\includegraphics[width=2.5in]{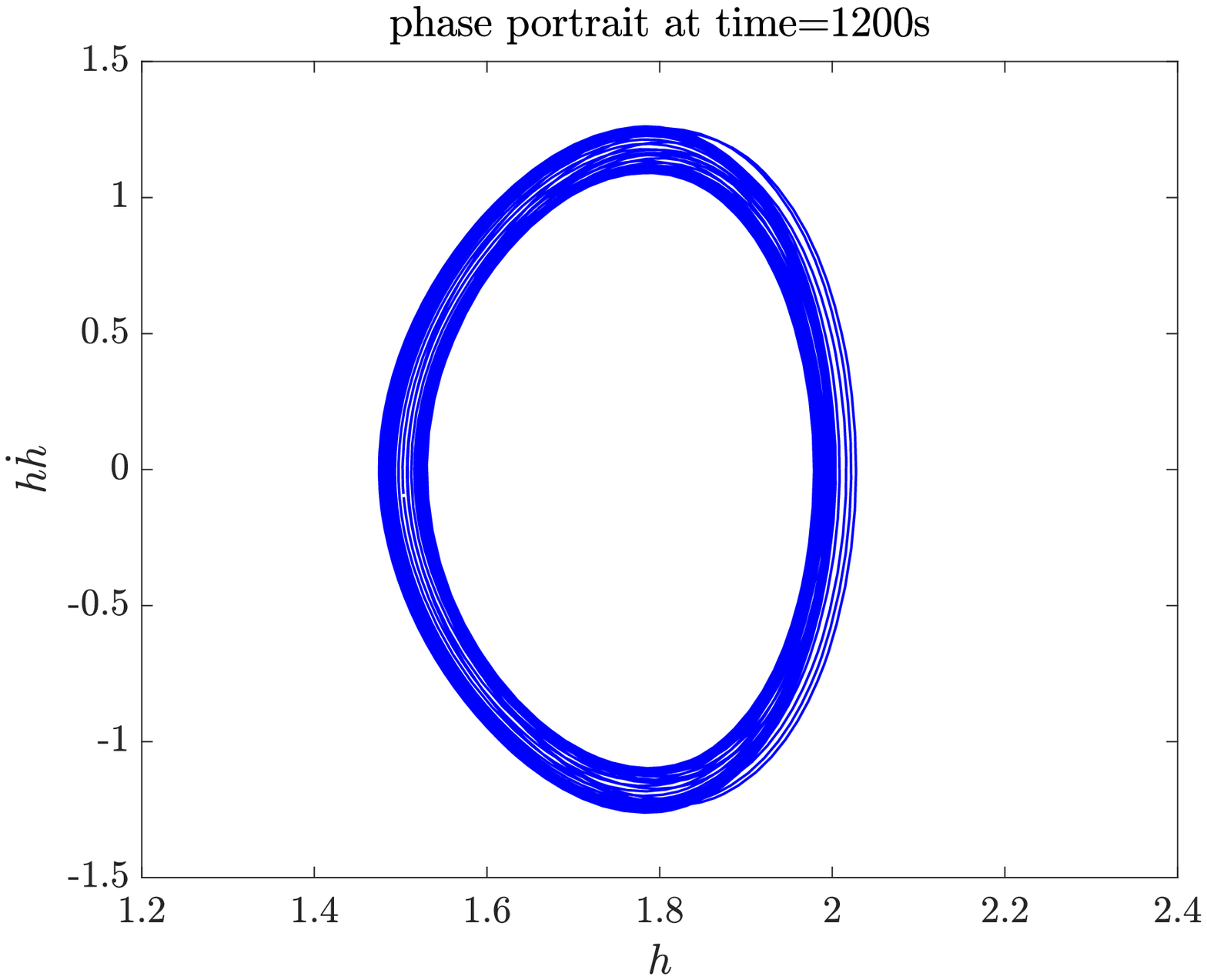}
\end{minipage}
\end{tabular}

\vskip 0.1in
\begin{tabular}{cc}
\begin{minipage}{2.6in}
\hskip 0.1in
\includegraphics[width=2.3in]{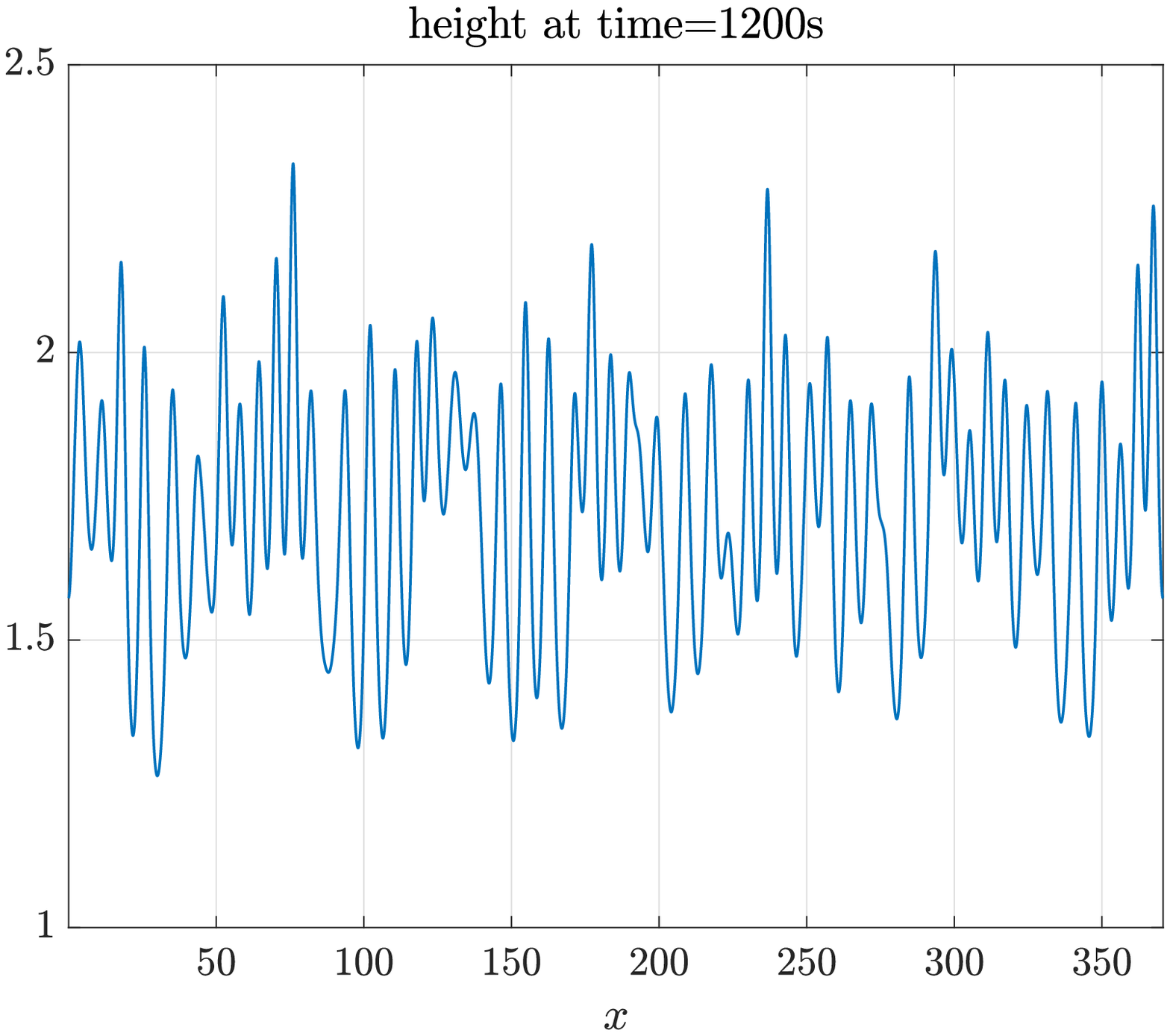}
\end{minipage}
&
\begin{minipage}{2.6in}
\includegraphics[width=2.5in]{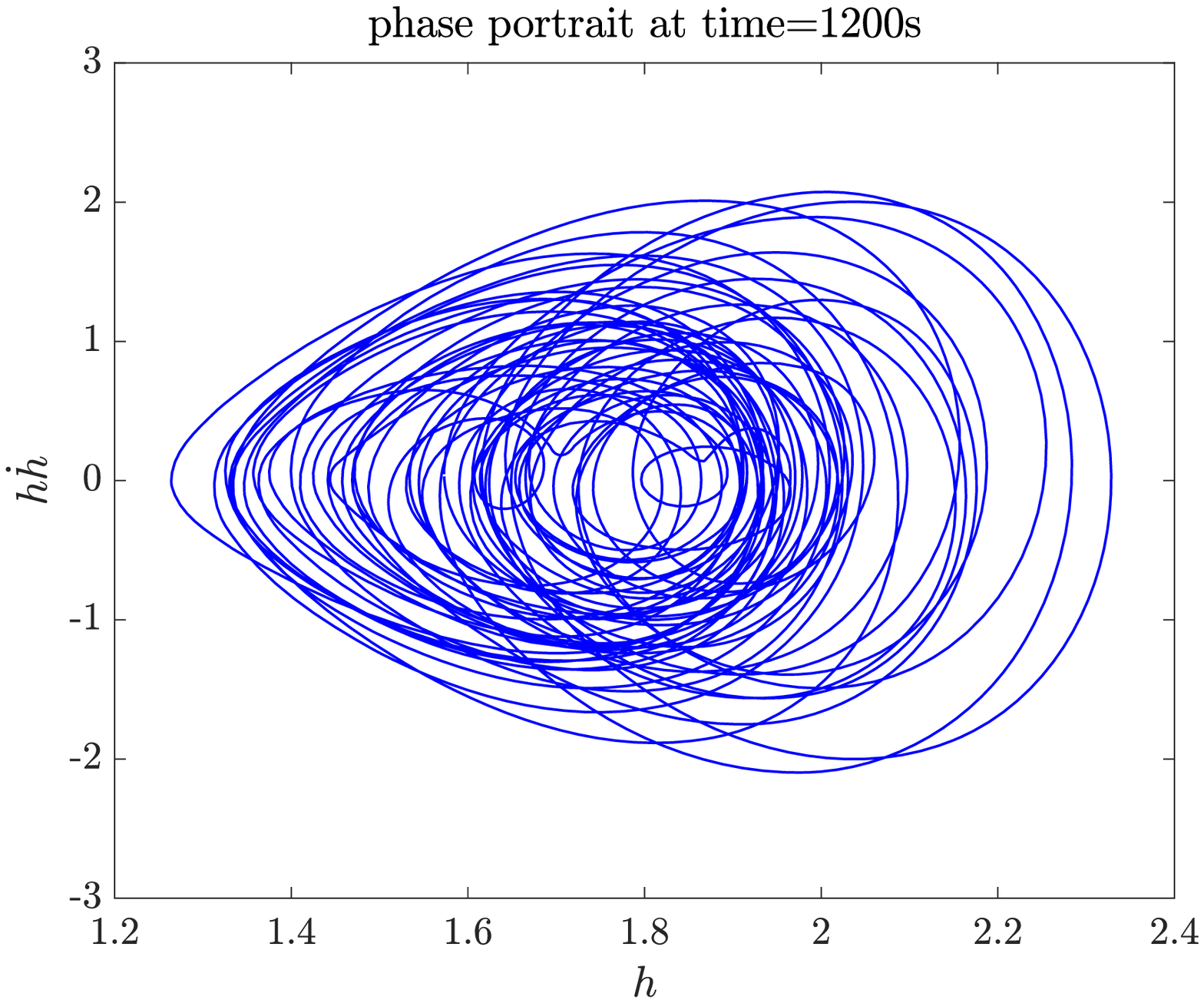}
\end{minipage}

\end{tabular}
\end{center}
\addtocounter{figure}{-1}
\vskip -0.1in
\caption{
     Continued.
         }
\end{figure}
\pagebreak
\section{Conclusions and perspectives}

We have  derived  the  modulation equations system to the SGN model and 
show that it is strictly hyperbolic for arbitrary  wave amplitudes, i.e. the periodic wave trains 
are modulationally stable.   This corroborates the results~\cite{Li_2001} where the linear stability 
of solitary waves (which can be considered as the limit of periodic waves of large length) 
has been proven. 
The existence of the Riemann invariants and nature of the characteristic fields 
(genuinely degenerate or genuinely nonlinear in the sense of Lax) will be the topic  for future research.

\setcounter{section}{0}
\renewcommand{\thesection}{\Alph{section}}

\section{Multiscale decomposition}\label{appendix_multiscale}

The classical Whitham method~\cite{Whitham_1974} consists in decomposing the scales in the 
following way (for simplicity, we consider just the velocity variable $u$):
\begin{equation}
 \label{decomposition_classical}
   u(x,t) = u\big(\theta (x,t),\varepsilon x,\varepsilon t\big), 
     \quad \theta (x,t)=\frac{\Theta(\varepsilon x, \varepsilon t)}{\varepsilon}.
\end{equation}
Here $\theta$ is a fast phase variable, $\Theta$ is a slow phase variable,  
and $\varepsilon$ is a small parameter. The solution is supposed to be $ 2\pi $-periodic 
with respect to $\theta$. The definitions of the local wave frequency $\omega$ 
and the local wave number $\kappa$ :
\begin{equation}\label{consistency_condition_fast}
	\pd{\theta}{t} = - \omega,\quad \pd{\theta}{x} = \kappa,
\end{equation}
automatically imply the evolution equation for $\kappa$:
\begin{equation}
 \label{consistency_equation_fast}
	\kappa_t+\omega_x = 0.
\end{equation}
Written in slow variables  $ X = \varepsilon x $, $ T = \varepsilon t $, 
equations~\eqref{consistency_condition_fast} are equivalent to
\begin{equation*}
  \pd{\Theta}{T} = - \omega,\quad \pd{\Theta}{X} = \kappa,
\end{equation*}
and~\eqref{consistency_equation_fast} reads as 
\begin{equation}
 \label{consistency_equation_slow}
	\kappa_T+\omega_X = 0.
\end{equation}
One can also define the travelling wave coordinate $ \xi = x - Dt $, 
and the phase  velocity  $ D = \omega/\kappa $. 
The solution is decomposed as:
\begin{equation}
 \label{multiscale_xi}
  u(x,t) = u(x-Dt,\varepsilon x,\varepsilon t) = u\Big(\frac{\theta(x,t)}{\kappa},
   \varepsilon x,\varepsilon t\Big) =  u\left(\frac{\Theta(X,T)}{\varepsilon \kappa},X,T\right).
\end{equation}
The wavelength $ L $ is defined as:
\begin{equation*}
   L = \frac{2\pi}{\kappa}.
\end{equation*}
Thus, since $ \theta \in [0,2\pi] $, 
then $ \xi = \frac{\Theta}{\varepsilon \kappa} \in \left[0,\frac{2\pi}{\kappa}\right] =[0,L]$. 
So, $u$ is an $L$-periodic function with respect to the travelling coordinate $\xi$:
\begin{equation*}
  u(\xi+L,X,T) = u(\xi,X,T).
\end{equation*}
Finally,~\eqref{consistency_equation_slow} reads:
\begin{equation}
 \label{wave_length_equation}
	(1/L)_T + (D/L)_X = 0.
\end{equation}
The approach employing the travelling coordinate  is equivalent to the one using the phase variable. 
The consistency equation can always be written in any of the two forms presented 
above:~\eqref{consistency_equation_slow} or~\eqref{wave_length_equation}. 

\section{Computation of elliptic integrals}
\label{appendix_definitions}
Let 
\begin{equation}
  h_2>h_1>h_0, \quad P_3(h)=(h-h_0)(h-h_1)(h_2-h).
\end{equation}
Then one has :
\begin{equation*}
 \begin{aligned}
  \frac{1}{2}\int_{h_1}^{h_2}\frac{h\,dh}{\sqrt{P_3(h)}} & =
   \sqrt{h_2-h_0}E(k)+\frac{h_0\, K(k)}{\sqrt{h_2-h_0}}, \\
 \frac{1}{2}\int_{h_1}^{h_2}\frac{dh}{\sqrt{P_3(h)}} & = \frac{K(k)}{\sqrt{h_2-h_0}}, \\
  \frac{1}{2}\int_{h_1}^{h_2}\frac{h^{-1}dh}{\sqrt{P_3(h)}} & =\frac{\Pi(n,k)}{h_2\sqrt{h_2-h_0}}. 
 \end{aligned}
\end{equation*}
Here $K(k)$, $ E(k) $ and $ \Pi(n,k) $ are the complete elliptic integrals of the first, second and third type, respectively :
\begin{equation*}
 \begin{aligned}
  K(k) & = \int_{0}^{\frac{\pi}{2}}\frac{d\theta}{\sqrt{1-k^2\sin^2\theta}},\\
  E(k) & =  \int_{0}^{\frac{\pi}{2}}\sqrt{1-k^2\sin^2\theta}d\theta,\\
 \Pi(n,k) & =  \int_{0}^{\frac{\pi}{2}}\frac{d\theta}{(1-n\sin^2\theta)\sqrt{1-k^2\sin^2\theta}}.
 \end{aligned}
\end{equation*}
The characteristic $n$ and  elliptic modulus $k$ are defined as :
\begin{equation*}
  n =\frac{h_2-h_1}{h_2}, \quad k^2=\frac{h_2-h_1}{h_2-h_0}.
\end{equation*}
The definition of averaging~\eqref{definition_averaging} implies then :
\begin{equation*}
 \begin{aligned}
	\overline{h}      & = h_0 + (h_2-h_0)\frac{E(k)}{K(k)},\\
	\overline{h^{-1}} & = \frac{\Pi(n,k)}{h_2K(k)},\\
	L                 & = \frac{4 \sqrt{I_3}}{\sqrt{3}}  \frac{K(k)}{\sqrt{h_2-h_0}}.
 \end{aligned}
\end{equation*}
The derivatives of the complete elliptic integrals read as follows:
\begin{equation*}
 \begin{aligned}
 \frac{d}{dk}K(k) &= \frac{1}{k(1-k^2)}E(k)-\frac{1}{k}K(k),\\
 \frac{d}{dk}E(k) &= \frac{1}{k}E(k) - \frac{1}{k}K(k),\\
 \frac{\partial}{\partial n}\Pi(n,k)&=-\frac{1}{2(1-n)(k^2-n)}E(k)-\frac{1}{2n(1-n)}K(k)+
               \frac{(k^2-n^2)}{2n(k^2-n)(1-n)}\Pi(n,k),\\
 \frac{\partial}{\partial k}\Pi(n,k)&=\frac{k}{(k^2-n)(1-k^2)}E(k)-\frac{k}{k^2-n}\Pi(n,k).
 \end{aligned}
\end{equation*}
Or, in terms of $h_i$:
\begin{equation*}
 \begin{aligned}
 \left.\left(\frac{d}{dk}K(k)\right)\right|_{k=\sqrt{\frac{h_2-h_1}{h_2-h_0}}} & =
    \frac{h_2-h_0}{h_1-h_0}\sqrt{\frac{h_2-h_0}{h_2-h_1}}E(k)-\sqrt{\frac{h_2-h_0}{h_2-h_1}}K(k),\\
  \left.\left(\frac{d}{dk}E(k)\right)\right|_{k=\sqrt{\frac{h_2-h_1}{h_2-h_0}}} & =
   \sqrt{\frac{h_2-h_0}{h_2-h_1}}E(k)-\sqrt{\frac{h_2-h_0}{h_2-h_1}}K(k),\\
  \left.\left(\frac{\partial}{\partial n}\Pi(n,k)\right)\right|_{k=\sqrt{\frac{h_2-h_1}{h_2-h_0}},\, 
   n=\frac{h_2-h_1}{h_2}}&=-\frac{h_2^2(h_2-h_0)}{2h_0h_1(h_2-h_1)}E(k) - \frac{h_2^2}{2h_1(h_2-h_1)}K(k) + 
       \\
      & \hskip 1.0in
      \frac{h_2(h_0h_2+h_1h_2-h_0h_1)}{2h_0h_1(h_2-h_1)}\Pi(n,k),\\
   \left.\left(\frac{\partial}{\partial k}\Pi(n,k)\right)\right|_{k=\sqrt{\frac{h_2-h_1}{h_2-h_0}},\, 
      n=\frac{h_2-h_1}{h_2}}&=\frac{h_2(h_2-h_0)}{h_0(h_1-h_0)}
       \sqrt{\frac{h_2-h_0}{h_2-h_1}}E(k) - \frac{h_2}{h_0}\sqrt{\frac{h_2-h_0}{h_2-h_1}}\Pi(n,k).
	\end{aligned}
\end{equation*}

\section{Differentials of $ \overline{h} $, $ \overline{h^{-1}}$ and $ L $}
	\label{appendix_differentials}
The differentials of $ n $ and $ k $ are given by:
\begin{equation*}
 \begin{aligned}
	 dn & = - \frac{1}{h_2}dh_1 + \frac{h_1}{h_2^2}dh_2,\\
	 dk & = \frac{1}{2(h_2-h_0)}\sqrt{\frac{h_2-h_1}{h_2-h_0}}dh_0 - 
               \frac{1}{2\sqrt{(h_2-h_1)(h_2-h_0)}}dh_1 + \frac{h_1-h_0}{2(h_2-h_0)
                 \sqrt{(h_2-h_1)(h_2-h_0)}}dh_2.
 \end{aligned}
\end{equation*}
Then, the differentials of $\overline{h}$ and $\overline{h^{-1}}$ are :
\begin{equation*}
 \begin{aligned}[c]
   d\overline{h} & = \left(\frac{1}{2} - \frac{h_2-h_0}{2(h_1-h_0)} \frac{E^2(k)}{K^2(k)}\right) dh_0 + 
                     \left(\frac{h_2-h_0}{2(h_2-h_1)} - \frac{h_2-h_0}{h_2-h_1}\frac{E(k)}{K(k)} + 
                     \frac{(h_2-h_0)^2}{2(h_2-h_1)(h_1-h_0)}\frac{E^2(k)}{K^2(k)}\right)dh_1 + \\
	&\quad
         \left(-\frac{h_1-h_0}{2(h_2-h_1)}+\frac{h_2-h_0}{h_2-h_1}\frac{E(k)}{K(k)} - 
           \frac{h_2-h_0}{2(h_2-h_1)}\frac{E^2(k)}{K^2(k)}\right)dh_2,\\
  d\overline{h^{-1}} &= \left(\frac{1}{2h_0(h_1-h_0)}\frac{E(k)}{K(k)}-
          \frac{1}{2h_0h_2}\frac{\Pi(n,k)}{K(k)} - \frac{1}{2h_2(h_1-h_0)}
          \frac{\Pi(n,k)E(k)}{K^2(k)}\right)dh_0 + 
         \left ( \frac{1}{2h_1(h_2-h_1)} - \right . \\
	&\quad
          \left . \frac{h_2-h_0}{2h_1(h_2-h_1)(h_1-h_0)}
          \frac{E(k)}{K(k)} - \frac{1}{2h_1(h_2-h_1)}\frac{\Pi(n,k)}{K(k)}+
           \frac{h_2-h_0}{2h_2(h_2-h_1)(h_1-h_0)}\frac{\Pi(n,k)E(k)}{K^2(k)}\right)dh_1 + \\
	&\quad 
         \left(-\frac{1}{2h_2(h_2-h_1)}+\frac{1}{2h_2(h_2-h_1)}\frac{E(k)}{K(k)} + 
           \frac{h_1}{2h_2^2(h_2-h_1)}\frac{\Pi(n,k)}{K(k)} - \frac{1}{2h_2(h_2-h_1)}
            \frac{\Pi(n,k)E(k)}{K^2(k)}\right)dh_2.
\end{aligned}
\end{equation*}
The differential of  $L$ is : 
\begin{equation*}
 \begin{aligned}
   dL & = \frac{2}{\sqrt{3}}\left(\frac{\sqrt{h_0h_1h_2}}{(h_1-h_0)\sqrt{h_2-h_0}}E(k)+
          \frac{h_1h_2}{\sqrt{h_2-h_0}\sqrt{h_0h_1h_2}}K(k)\right)dh_0 + \\
	&\qquad
       \frac{2}{\sqrt{3}}\left(-\frac{\sqrt{h_2-h_0}\sqrt{h_0h_1h_2}}{(h_2-h_1)(h_1-h_0)}E(k)+
          \frac{h_0h_2^2}{(h_2-h_1)\sqrt{h_2-h_0}\sqrt{h_0h_1h_2}}K(k)\right)dh_1 + \\
	&\qquad\frac{2}{\sqrt{3}}\left(\frac{\sqrt{h_0h_1h_2}}{(h_2-h_1)
           \sqrt{h_2-h_0}}E(k)-\frac{h_0h_1^2}{(h_2-h_1)\sqrt{h_2-h_0}\sqrt{h_0h_1h_2}}K(k)\right)dh_2.
 \end{aligned}
\end{equation*}
	
{\bf Acknowledgments} Part of this work  was performed  during the  visit of SG  to 
the Department of Mathematics of  National Taiwan University. 
SG was partially supported by l'Agence Nationale de la Recherche, France
(grant numbers ANR-11-LABEX-0092, and ANR-11-IDEX-0001-02).  
KMS was partially supported by MOST 105-2115-M-002-013-MY2
and 107-2115-M-002-014.

\pagebreak
	
\end{document}